\documentclass [aps,prb,twocolumn,amssymb,amsmath,showpacs,floatfix]{revtex4-2}

\makeatletter
  \long\def\pprintMaketitle{\clearpage
  \iflongmktitle\if@twocolumn\let\columnwidth=\textwidth\fi\fi
  \resetTitleCounters
  \def\baselinestretch{1}%
  \printFirstPageNotes
  \begin{center}%
 \thispagestyle{pprintTitle}%
   \def\baselinestretch{1}%
    \Large\@title\par\vskip18pt
    \normalsize\elsauthors\par\vskip10pt
    \footnotesize\itshape\elsaddress\par\vskip36pt
    \end{center}%
  \gdef\thefootnote{\arabic{footnote}}%
  }
\makeatother

\usepackage{amsmath}
\usepackage{amssymb}
\usepackage{amsfonts}
\usepackage{graphicx}
\usepackage{verbatim}
\usepackage{bm}
\usepackage{mathbbol}
\usepackage{color}
\setcitestyle{compress}

\usepackage{mathtools}

\usepackage{enumitem}

\usepackage{graphicx,epsfig,amsfonts,amssymb}
\usepackage{bm}
\usepackage{times}
\usepackage{lipsum}
\usepackage{verbatim}

\usepackage{hyperref}
\hypersetup{
    colorlinks,
    citecolor=blue,
    filecolor=blue,
    linkcolor=blue,
    urlcolor=blue
}

\newcommand{\beg}{\begin{equation}}
\newcommand{\en}{\end{equation}}

 \newcommand{\lam}{\lambda}

\newcommand{\eref}[1]{Eq.~(\ref{#1})}
\newcommand{\re}[1]{(\ref{#1})}

\newcommand{\esref}[1]{Eqs.~(\ref{#1})}
\renewcommand{\Re}{\mathrm{Re}}
\renewcommand{\Im}{\mathrm{Im}}
\newcommand{\Tr}{\mathrm{Tr}\,}
\newcommand{\eps}{\varepsilon}

\newcommand{\dn}{\downarrow}
\newcommand{\up}{\uparrow}

\newcommand{\rr}{{\bm r}}
\newcommand{\pp}{{\bm p}}
\newcommand{\qq}{{\bm q}}

\newcommand{\ii}{{\bm i}}
\newcommand{\jj}{{\bm j}}
\newcommand{\ps}{\mathsf{p}}
\newcommand{\qs}{\mathsf{q}}

\newcommand{\GG}{\mathcal{G}}
\newcommand{\FF}{\mathcal{F}}

\renewcommand{\iint}{\int\!\!\!\int}

 \newcommand{\overbar}[1]{\mkern 1.5mu\overline{\mkern-1.5mu#1\mkern-1.5mu}\mkern 1.5mu}

\makeatletter
\renewcommand{\p@subsection}{\thesection }
\makeatother
  
\begin{document}

\title{Migdal-Eliashberg theory as a classical spin chain}

\author{Emil A. Yuzbashyan}
\affiliation{Department of Physics and Astronomy, Rutgers University, Piscataway, New Jersey 08854, USA}

\author{Boris L. Altshuler}
\affiliation{Physics Department, Columbia University, 538 West 120th Street, New York, New York 10027, USA}

 \begin{abstract}
  We formulate the Migdal-Eliashberg theory of electron-phonon interactions in terms of classical spins by mapping  the free energy    to a Heisenberg spin chain in a Zeeman magnetic field. Spin components  are energy-integrated normal and   anomalous Green's functions and sites of the chain are fermionic Matsubara frequencies. The Zeeman field grows linearly with the spin coordinate and competes with ferromagnetic spin-spin interaction  that falls off as the square of the inverse distance.  The spin-chain representation makes a range of previously unknown properties plain to see. In particular,  infinitely many new solutions of the Eliashberg equations both in the normal and superconducting states  emerge at strong coupling. These  saddle points of the free-energy functional  correspond to spin flips. We argue that they are also fixed points of kinetic equations and play an essential role in   far from equilibrium dynamics of strongly coupled superconductors. Up to an overall phase, the frequency-dependent gap function that minimizes the free energy \textit{must} be non-negative. There are strong parallels between  our \textit{Eliashberg spins} and Anderson pseudospins,  though the two sets of spins never coincide.

 \end{abstract}
 
\maketitle

\section{Introduction}

Electron-phonon interactions determine  many properties of quantum metals, such as  the  charge and heat transport, thermodynamics, and  superconductivity. A well-established approach to these phenomena is the Migdal-Eliashberg   theory~\cite{migdal, eli1st}. This theory operates with two functions of   Matsubara frequency, $\Sigma_n\equiv \Sigma(\omega_n)$ and $\Phi_n\equiv\Phi(\omega_n)$,  which have the meaning of the normal and anomalous self-energies.  The   functions $\Sigma_n$ and $\Phi_n$ must satisfy  two nonlinear algebraic equations  known as the \textit{Eliashberg equations}, whose kernel   is the  phonon-mediated electron-electron interaction $\lam_l\equiv\lam(\omega_l)$. The order parameter   is the frequency-dependent gap function $\Delta_n=\Phi_n/Z_n$, where $Z_n=1+\Sigma_n/\omega_n$.

The main dimensionless parameter in the Migdal-Eliashberg theory is the  electron-phonon coupling defined as  the electron-electron interaction evaluated at zero Matsubara frequency, $\lam=\lam(\omega_l=0)$. The coupling constant $\lam$ is inversely proportional to the square of the characteristic  phonon frequency $\omega_\mathrm{ch}$. In conventional electron-phonon models, such as the Holstein or Fr\"olich Hamiltonian,  interactions with electrons strongly renormalize   phonon frequencies~\cite{migdal,eli1st,agd}.  Because of this it is crucial to distinguish \textit{bare} ($\lam_0$) and \textit{renormalized} ($\lam$) coupling constants~\cite{brauer}.

 The Migdal-Eliashberg theory enjoyed a great deal of success over many decades as a quantitative theory of conventional
superconductivity.
With  phonon spectrum extracted  from experiment, the   theory makes accurate predictions for the superconducting transition temperature $T_c$,  zero temperature energy gap, jump in the specific heat at $T_c$, density of states, etc., for a broad range of superconductors, such as Al, V, Ta, Sn,  Tl, In, Nb, Pb, their various alloys, Hg, and MgB$_2$~\cite{allendynes,mitrovic,carbotte,frank,choi,mcmillan}. The  coupling constant   in these materials ranges between $\lam\approx0.4$ for Al and $\lam\approx 1.6$ for Hg in simple elements and up
to $\lam\approx 3.0$ (Pb$_{0.5}$Bi$_{0.5}$) in alloys.

Despite the success and maturity of the Migdal-Eliashberg theory, there is also a great deal of controversy surrounding it. 
A number of publications~\cite{roland,alexandrov,meyer,capone,esterlis,scalapino}  rediscover   Migdal and Eliashberg's observation~\cite{migdal, eli1st} that the  theory is inapplicable to $\lam_0\gtrsim 0.5,$    Migdal theorem notwithstanding~\cite{migdalthm}. Others see no upper limit on $\lam$~\cite{brauer} and even hypothesize that the strong coupling, $\lam\to\infty$, limit of the Migdal-Eliashberg theory  is attainable~\cite{andrey_validity}. A recent study claims   Eliashberg equations acquire a one-parameter family of solutions in the strong coupling limit  and as a consequence the superconducting transition temperature vanishes~\cite{andreygamma2}. Even the   $\lam\to0$ limit   is not free from controversy. While it is generally accepted that the weak coupling limit of the Migdal-Eliashberg theory is the more famous Bardeen-Cooper-Schrieffer (BCS) theory~\cite{bcs}, some argue that this is in fact untrue~\cite{mars1,mars2}.

Our goal is to resolve these issues in a series of  papers~\cite{retardation,meaningmigdal,breakEli}. Here we establish several important properties of the free energy of the electron-phonon system.  We show that the   gap function $\Delta_n$ must be non-negative   at the global minimum point and argue   that   the   minimum is unique by symmetry    at all temperatures and coupling strengths, up to an overall phase $e^{i\phi}$.  
  Aside from the global minimum,   we find an infinite set of new saddle points  at strong coupling.
  As usual, stationary points other than the global minimum do not affect  equilibrium properties in the thermodynamic limit. However,   we reason that these  saddle points play a major role in the far from equilibrium dynamics of the electronic subsystem. Besides, their emergence
  and proliferation  seem to be tied to the subsequent breakdown of the Migdal-Eliashberg theory discussed in the next paragraph.
   For the sake of completeness, we  also provide an educational appendix where we explain within the path integral framework
that the BCS theory undoubtedly \textit{is} the  weak coupling limit of the Migdal-Eliashberg theory when this limit is properly taken.

In the next paper~\cite{breakEli},  we will  show that the Migdal-Eliashberg theory loses its validity at a \textit{finite}   value of the \textit{renormalized} electron-phonon coupling $\lam_c$ due to a  phase transition  which breaks the lattice translational symmetry. We will then construct a new theory  (theory of lattice-fermionic superfluidity) that  works past $\lam_c$.  Prior to the transition it reduces to the Migdal-Eliashberg theory. Afterwards, it describes the new state of the system. 

\begin{figure}
	\centering
	 \includegraphics[width=0.9\columnwidth]{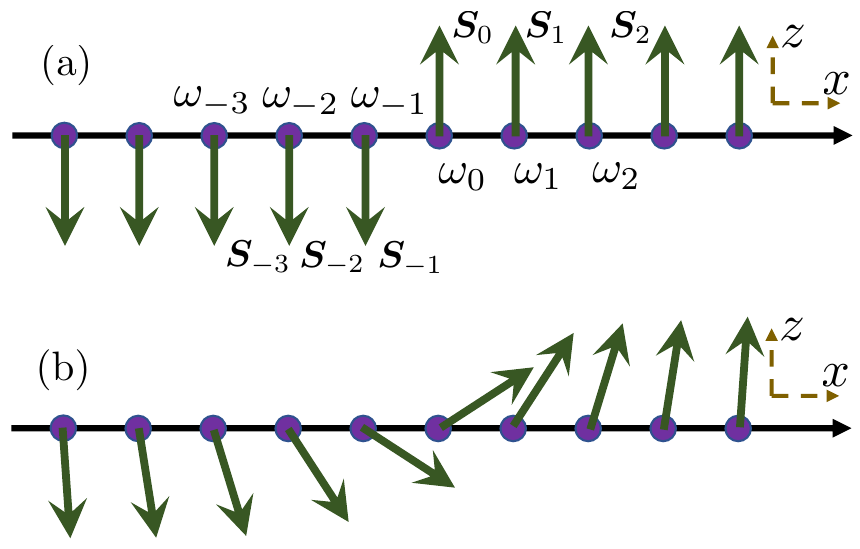}
	\caption{Superconducting transition in terms of classical spins. As shown in the text,  Migdal-Eliashberg theory maps to a classical Heisenberg spin chain. The positions of the spins are fermionic Matsubara frequencies $\omega_n$. The spin-spin interactions are purely ferromagnetic and the spins are subject to a Zeeman magnetic field $2\pi\omega_n$  along the $z$ axis.  The figure
	shows (a) the normal state and (b) a superconducting state.  In the superconducting state, spins acquire  $x$ components, which implies nonzero anomalous  Green's function. The sharp domain wall in the normal state is smeared in the superconducting state. The normal-superconductor transition  therefore translates into softening of the domain wall  in the spin language.}
	\label{spinsfig}
\end{figure}

Our main tool to achieve these goals is a representation of the Migdal-Eliashberg theory as a classical Heisenberg spin chain, which we describe below. Positions of the \textit{Eliashberg spins} are the fermionic Matsubara frequencies $\omega_n=\pi T(2n+1)$.  Spin components are momentum-integrated normal and anomalous Green's functions. The interaction between them is ferromagnetic  and falls off as $(\omega_n-\omega_m)^{-2}$
at large  separation. In addition, the spins are  subject to a position-dependent Zeeman field $2\pi\omega_n$ along the $z$ axis. The   energy of the spin chain is proportional to the free-energy density   of the electron-phonon system. 
Classical spins provide a simple and intuitive picture of the normal and superconducting states and of the transition between them.
 In the normal state, the spins are parallel to the $z$ axis, $\bm S_n=\mbox{sgn} (\omega_n) \hat {\bm z}$, where $\hat {\bm z}$ is a unit vector along the $z$ axis. Below the superconducting transition temperature, they acquire $xy$  components softening the sharp domain wall between $\omega_{-1}$ and $\omega_0$ in the normal state as shown in Fig.~\ref{spinsfig}. 
 
 Not surprisingly, Eliashberg equations emerge as stationary point equations for the free-energy functional. Their solutions   therefore correspond to either minima or maxima or saddle points of the free energy.
The spin-chain formulation makes it straightforward to demonstrate several fundamental  properties  of these stationary points.  In particular, we use it to prove that   the Eliashberg gap  function at the global minimum is of the form $\Delta_n=e^{i\phi}|\Delta_n|$, where $|\Delta_n|$ is   even in  $\omega_n$.   Furthermore, an infinite discrete set of saddle points emerges as we increase $\lam$.  
In terms of the  spin chain, they  are  equilibria with a certain number of spins flipped against  effective magnetic fields acting on them.   The minimum is  the stable, lowest energy spin configuration.  The new saddle points begin to proliferate  just before the Migdal-Eliashberg theory breaks down. We argue that these saddle points play a special role in the far from equilibrium collisionless dynamics
of strongly coupled conventional superconductors. Namely, they are the fixed points of the corresponding kinetic equations, which are Hamilton's equations of motion for Eliashberg spins. When sufficiently many spins are flipped, these stationary points are unstable and
give rise to rich soliton-like dynamics.

The content of this paper is as follows. In Sec.~\ref{chainsect} we map the Migdal-Eliashberg theory to a classical spin chain building on the path integral formulation we develop in Appendixes~\ref{holsteinapp} and \ref{standardapp}.  We interpret the superconducting transition in terms of spins in Sec.~\ref{spinsect}, construct a divergence-free form of Eliashberg equations as well as new (spin-flip) solutions for them
in Sec.~\ref{newsect}, and   relate spins and the  Eliashberg gap function $\Delta_n$ in Sec.~\ref{gapsect}. 
 These three sections lay the groundwork for the analysis of  stationary points of the free energy in Sec.~\ref{stptsec}. Section~\ref{strongsect} focuses on the strong coupling limit $\lam=\infty$ and in Sec.~\ref{ander} we compare the two notions of spins in the theory of superconductivity (Eliashberg spins introduced in this paper and Anderson pseudospins~\cite{pseudo}) and discuss the role of the spin-flip solutions of the Eliashberg equations in the collisionless dynamics of strongly coupled superconductors.

\section{Mapping  to a Heisenberg spin chain}
\label{chainsect}

We work with two models in this paper. One is the Holstein model (dispersionless phonons) with  arbitrary hopping matrix and onsite potential. The other is  a rather general electron-phonon Hamiltonian with arbitrary phonon dispersion and momentum-dependent electron-phonon coupling. 
The Holstein Hamiltonian reads as
\beg
H=\sum_{\bm i \bm j \sigma} h_{\bm i\bm j} c^\dagger_{\bm i\sigma} c_{\bm j \sigma}+  \sum_{\bm i} \left[\frac{p_\ii^2}{2M} + \frac{K_0 x_\ii^2}{2}\right] +
\alpha \sum_{\bm i} n_{\bm i} x_\ii,
\label{holsteinH}
\en 
where $\ii$ and $\jj$ label  lattice sites, $h_{\ii\jj}$ are the matrix elements of an \textit{arbitrary} single-electron Hamiltonian $\hat h$~\cite{arbitrary}, $c_{\bm i \sigma}$   annihilates an electron on site $\ii$ with spin projection $\sigma$,  $n_\ii =\sum_\sigma c^\dagger_{\ii\sigma} c_{\ii\sigma}$ is the fermion occupation operator, and $p_\ii$ and $x_\ii$ are ion momentum and position operators. The \textit{bare} phonon frequency is $\Omega_0=\sqrt{K_0/M}$. The lattice and its dimensionality are at this point arbitrary.

 The   Hamiltonian for electrons interacting with dispersing phonons is
\beg
\begin{split}
H=&\sum_{\pp \sigma} \xi_\pp c^\dagger_{\pp\sigma} c_{\pp\sigma} +  \sum_\qq {\omega_{0}(\qq) } b^\dagger_\qq b_\qq+\\
&+\frac{1}{\sqrt{N}} \sum_{\pp \qq\sigma} \frac{\alpha_{\qq}}{\sqrt{2M\omega_0(\qq)}} c^\dagger_{\pp+\qq \sigma} c_{\pp\sigma} \left[ b^\dagger_{-\qq} + b_\qq\right],
\end{split}
\label{frol1}
\en
where $M$ is the ion mass and $N$ is the number of lattice sites.
 
We proceed with a path-integral formulation of the Migdal-Eliashberg theory. The first step is to integrate out  phonons, which leaves us with an effective fermion-fermion interaction quartic in  fermionic fields. The second step is to decouple the quartic term with three Hubbard-Stratonovich fields
$\Phi$, $\Sigma_\up$, and $\Sigma_\dn$ that are functions of two imaginary times and, in general, two space points. In the  next step, we integrate out the fermions to obtain an effective action solely in terms of the \textit{Eliashberg fields} $\Phi$, $\Sigma_\up$, and $\Sigma_\dn$. The fourth and  final step is to obtain the stationary point of this effective action, where the Eliashberg fields depend on the time difference only. We detailed the above procedure  in Appendix~\ref{holsteinapp} for
the Holstein model and  in Appendix~\ref{standardapp} for dispersing phonons~\re{frol1}. The end result is an expression for the free-energy functional~\cite{grand} of the system per site,
\beg
\begin{split}
f= \nu_0 T^2\sum_{nl}\left[ \Phi_{n+l}^* \Lambda_l \Phi_{n} + \Sigma_{n+l} \Lambda_l  \Sigma_{n}    \right]-\\
2\pi \nu_0 T\sum_{n} \sqrt{ (\omega_n+\Sigma_{n} )^2 +| \Phi_{n} |^2 },
\end{split}
\label{feH1}
\en
where $\nu_0$ is the density of states at the Fermi energy per site per spin projection. The  field $\Phi_n\equiv\Phi(\omega_n)$ is complex and $\Sigma_n\equiv\Sigma(\omega_n)$ is real. Both fields are functions of the fermionic Matsubara frequency $\omega_n$. At the stationary point, these fields equal the anomalous and normal self-energies. 

The quantity $\Lambda_l$ in \eref{feH1} is the Fourier transform of $1/\lambda(\tau)$ at bosonic Matsubara
frequency $\omega_l=2\pi T l$. Here $\lam(\tau)$ is the effective fermion-fermion interaction in the imaginary-time domain. 
 In the Matsubara frequency  domain this interaction for the Holstein model reads as
\beg
\lambda(\omega_l)= \frac{  g^2}{\omega_l^2+\Omega^2},\quad g^2=\nu_0 \alpha^2 M^{-1}.
\label{Holstein_lambda}
\en  
 We also define the dimensionless electron-phonon coupling constant as
 \beg
 \begin{split}
 \lambda=\lambda(\omega_l=0)=\frac{g^2}{\Omega^2}=\frac{\nu_0\alpha^2}{K},  \\
 \end{split}
 \label{lam}
 \en
 where $\Omega$ and $K$ are the \textit{renormalized} phonon frequency and spring constant.  See Appendix~\ref{standardapp} for $\lam(\omega_l)$ and $\lam=\lam(\omega_l=0)$ for dispersing phonons. At times we will consider the strong coupling limit $\lam\to\infty$, which is equivalent to $\Omega\to 0$ or $K\to0$ for the Holstein model.
   
 The stationary point equations for the free energy~\re{feH1}, are the \textit{Eliashberg equations}~\cite{eli1st}
\begin{subequations}
\begin{eqnarray}
\Phi_n= \pi  T \sum_m  \lambda_{nm}\frac{ \Phi_m}{ \sqrt{(\omega_m+\Sigma_{m} )^2 +| \Phi_{m} |^2} },\label{Phieq} \\ 
\Sigma_n= \pi  T \sum_m  \lambda_{nm}  \frac{\omega_m+ \Sigma_m}{ \sqrt{(\omega_m+\Sigma_{m} )^2 +| \Phi_{m} |^2}}, \label{Sigmaeq}
\end{eqnarray}
\label{elifamiliar3}
\end{subequations}
where 
\beg
\lam_{nm}=\lambda(\omega_n-\omega_m).
\label{lambda_nm}
\en
Note also that 
\beg
\lambda_{nn}=\lambda
\label{lambda_nn}
\en
diverges in the strong coupling limit.

It is convenient to introduce new variables, complex $F(\tau)$ and real $G(\tau)$, such that
\beg
\Phi(\tau)=\pi \lambda(\tau) F(\tau),\quad \Sigma(\tau)=\pi \lambda(\tau) G(\tau).
\en
In frequency representation we have
\beg
\begin{split}
\Phi_n=\pi T\sum_m  \lambda_{nm} F_m,\\
 \Sigma_n=\pi T\sum_m  \lambda_{nm} G_m,
 \end{split}
 \label{varchange}
\en
and \eref{elifamiliar3} becomes
\beg
\begin{split}
 F_n=\frac{ \Phi_n}{ \sqrt{(\omega_n+\Sigma_{n} )^2 +| \Phi_{n} |^2} },\\
G_n = \frac{\omega_n+ \Sigma_n}{ \sqrt{(\omega_n+\Sigma_{n} )^2 +| \Phi_{n} |^2}}.
\end{split}
\label{fg}
\en
On the stationary point, the fields $F_n$ and $G_n$ correspond to the anomalous and normal Green's functions integrated over the single-particle energy (see Appendix~\ref{gfapp}). We also show in Appendix~\ref{gfapp} that assuming  time-reversal symmetry, $\Sigma_n$ is real and odd in $\omega_n$ and $|\Phi_n|$ is even. Therefore, $G_n$ is real and odd and $|F_n|$ is even.

Observe that \eref{fg} implies a constraint on the variables  $G_n$ and $F_n$:
\beg
G_n^2+|F_n|^2=1.
\label{constraint}
\en
This means that we can trade these variables for three components of a \textit{classical spin} $\bm S_n$ of unit length, $\bm S^2_n=1$:
\beg
S_n^z=G_n,\quad S_n^x=\Re (F_n),\quad S_n^y=\Im (F_n).
\label{sgf}
\en
It follows from \eref{fg} that
\beg
F_n\Phi_n^*+G_n(\omega_n +\Sigma_n)= \sqrt{(\omega_n+\Sigma_{n} )^2 +| \Phi_{n} |^2}.
\label{sqrt}
\en
This allows us to rewrite the free energy~\re{feH1} as
\beg
H_s\equiv\frac{f}{\nu_0 T}=-2\pi \sum_n \omega_n S_n^z-\pi^2 T\sum_{nm} \lambda_{nm}{\bm S_n}\cdot {\bm S_m}.
\label{spinh}
\en

We interpret the free energy~\re{spinh}  as a Hamiltonian, $H_s$, of an open classical Heisenberg spin chain in an inhomogeneous Zeeman magnetic field. The positions of the spins are fermionic Matsubara frequencies $\omega_n$. Spin-spin interactions are ferromagnetic and fall off at large ``distance" as $\lambda_{nm}\propto (\omega_n-\omega_m)^{-2}$. The ``magnetic field'' is linear in the position of the spin and goes to $\pm\infty$ as $\omega_n\to\pm\infty$. Note that the Boltzmann weight   is $e^{-f N/T}=e^{-\nu_0 N H_s}=e^{-H_s/\delta}$, where $\delta=(\nu_0 N)^{-1}$ is the single-particle (electron) level spacing in the original electron-phonon problem. Therefore, the spin chain is at an effective temperature
\beg
T_s=\delta.
\en
Let us also write down the classical spin Hamiltonian  for the Holstein model as a visual [substitute \eref{Holstein_lambda} into \eref{spinh}]:
\beg
H_s=-2\pi \sum_n \omega_n S_n^z-\pi^2 Tg^2\sum_{nm} \frac{ {\bm S_n}\cdot {\bm S_m}}{(\omega_n-\omega_m)^2+\Omega^2}.
\label{schol}
\en
See the list below \eref{gapeqmin} for more properties of the spin-chain representation of the free energy.

Sometimes an on-site Hubbard repulsion is added~\cite{allendynes} to the Eliashberg equations by replacing $\lambda_{nm}$
 in~\eref{Phieq} with
$\lambda_{nm}-u$. This adds a long-range $xy$  term to the spin Hamiltonian~\re{spinh},
\beg
H_C=\pi^2Tu\sum_{nm}   \left(S_n^xS_m^x+S_n^y S_m^y\right),
\en
and the classical spin Hamiltonian becomes $H_s+H_C$.

\section{Superconductivity in terms of spins}
\label{spinsect}

In the spin language, the superconducting transition translates into  softening of the domain wall as shown in Fig.~\ref{spinsfig}.  This is
similar to the Anderson pseudospin description of the BCS superconductivity~\cite{pseudo}   even though Eliashberg  and Anderson  spins are not the same (see Sec.~\ref{ander}).

To understand what happens  as we lower the temperature, let us analyze the spin texture minimizing the energy $H_s$ as a function of $T$.
The Zeeman magnetic field and ferromagnetic interaction compete in $H_s$. The spin configuration minimizing the  Zeeman term is $\bm S_n=\mbox{sgn} (\omega_n) \hat {\bm z}$,  and the Zeeman field  inevitably prevails far from the origin, so that $\bm S_n\to \pm \hat {\bm z}$ for $\omega_n\to\pm\infty$. These fixed values at infinity    serve as
boundary conditions for the interaction term.

Above the superconducting $T_c$, the anomalous averages vanish, $F_n=0$. According to the definition~\re{sgf} of the classical spin, this means that all spins are parallel to the $z$ axis. From the behavior of $\bm S_n$ at large $\omega_n$ and by symmetry, it is then clear that the minimum energy spin texture is $\bm S_n=\mbox{sgn} (\omega_n) \hat {\bm z}$. This is the \textit{normal state} in the spin language. The characteristic feature of the normal state is a sharp domain wall between $\omega_{-1}$ and $\omega_0$ with an abrupt maximal jump of the $z$-component of spin from $S_{-1}^z=-1$ to $S_0^z=+1$ (see Fig.~\ref{spinsfig}). 

Below $T_c$ the anomalous averages are nonzero, i.e.,  the spins acquire $x$ components ($F_n$ can be made real in the spin-chain ground state). This means softening of the domain wall. Now the change in $S_n^z$  from $-1$ at $\omega_n\to-\infty$ to $+1$ at $\omega_n\to+\infty$ is gradual and the jump $S_{0}^z-S_{-1}^z <2$. Spin configurations with nonzero $xy$  components are superconducting states.

\section{ New  solutions  of Eliashberg equations }
\label{newsect}

We begin this section by verifying that solutions of the
  Eliashberg equations are equilibrium points of the spin chain.   
  This correspondence  allows us to identify new solutions of the Eliashberg equations both with zero and nonzero anomalous self-energy $\Phi_n$, i.e., both normal and superconducting. These solutions correspond to equilibria of the spin chain with a number of spins flipped against their effective magnetic fields, as opposed to all spins being along their fields as in the ground state.
 
 \subsection{Eliashberg equations as equilibrium condition for spins}
     
 Consider our spin Hamiltonian in its most general form
 \beg
H_s =-2\pi \sum_n \omega_n S_n^z-\pi^2 T\sum_{nm} \lambda_{nm}{\bm S_n}\cdot {\bm S_m}.
\label{spinh21}
\en
Terms in the Hamiltonian that contain spin $\bm S_n$ are
 \beg
 h_n=-\bm b_n\cdot \bm S_n,\quad \bm b_n=2\pi\omega_n\hat {\bm z}+2\pi^2 T\sum_{m\ne n} \lam_{nm} \bm S_m,
 \label{bn}
 \en
 where $\bm b_n$  is the effective magnetic spin acting on spin $\bm S_n$, which is the Zeeman field plus the field from other spins.
 
 Equilibrium is when each spin  is collinear with its field (parallel or antiparallel), i.e.,
 \beg
 \bm S_n= e_n \frac{\bm b_n}{|\bm b_n|},\quad e_n=\pm1,
 \label{Sb}
 \en
 where $e_n=+1$ indicates that the spin $\bm S_n$ is parallel to its effective field $\bm b_n$, while $e_n=-1$, that the spin  is flipped (antiparallel to the field).
 However, not all spin equilibria correspond to stationary points of the Eliashberg free energy~\re{feH1}. The reason is that we extracted
 the square root in \eref{sqrt} to obtain the spin chain. Only such spin equilibria are stationary points of the free energy for which the 
 left-hand side of \eref{sqrt} is non-negative. 
 
 To interpret this condition in terms of spins, let us introduce an additional magnetic field, which includes the action of the spin on itself,
 \beg
 \bm B_n=\bm b_n +2\pi^2 T\lam\bm S_n.
 \label{Bdef}
 \en
 Recall that $\lam=\lam(\omega_l=0)=\lam_{nn}$. The definition of $\bm b_n$ in \eref{bn}, the definition of spins~\re{sgf}, and \eref{varchange} imply
 \beg
 B_n^+=2\pi\Phi_n,\quad B_n^z=2\pi(\omega_n+\Sigma_n).
 \label{bps}
 \en
 Here and below we use the notation $V^+\equiv V^x+iV^y$, where $V^x$ and $V^y$ are the $x$ and $y$ components of a vector 
 $\bm V$. Since in equilibrium $\bm b_n$ is collinear with $\bm S_n$, so is $\bm B_n$. Then, the requirement that the expression $F_n\Phi_n^*+G_n(\omega_n +\Sigma_n)$ on the left-hand side of \eref{sqrt} be non-negative is equivalent to
 \beg
 \bm B_n\cdot \bm S_n=\bm b_n\cdot \bm S_n+2\pi^2 \lam T \ge 0.
 \label{flipcond}
 \en
 This condition always holds when $e_n=+1$ for all $n$ (no spin flips). 
 
 We saw that in equilibrium $\bm S_n$ is  either parallel or antiparallel to $\bm B_n$. Equilibria  of the spin chain with each $\bm S_n$ parallel to $\bm B_n$ correspond to solutions of the Eliashberg equations and vice versa. We consider only this type of equilibria in this paper. However, it is important to keep in mind that spin flips are nevertheless  allowed as $\bm S_n$ can be parallel to $\bm B_n$, but antiparallel to $\bm b_n$. Later in this section we will point out equilibrium  spin configurations of this type. To derive Eliashberg equations from $\bm S_n\parallel \bm B_n$, note that this condition is equivalent to
 \beg
 S_n^z= \frac{B_n^z}{|\bm B_n|},\quad  S_n^+= \frac{B_n^+}{|\bm B_n|}.
 \label{SB}
 \en
 Equation~\re{Bdef} now implies the following two self-consistency conditions:
 \beg
 B_n^z= \sum_m \lam_{nm} \frac{B_m^z}{|\bm B_m|},\quad  B_n^+= \sum_m \lam_{nm} \frac{B_m^+}{|\bm B_m|}.
 \en
 Using \eref{bps}, we see that these are precisely the Eliashberg equations~\re{elifamiliar3}.

 The spin-chain approach naturally leads to an alternative set of Eliashberg equations that  are free of divergencies in the strong coupling limit $\lam\to\infty$ both in the superconducting and normal states~\cite{andrey_validity}. Indeed, we see from \esref{elifamiliar3} and \re{lambda_nn}
 that this limit presents a problem for the Eliashberg equations as they contain a diverging $\lam_{nn}=\lam$ term. As a result, both self-energies, $\Sigma_n$ and $\Phi_n$,  are proportional to $\lam$ and diverge in this limit. Equation~\re{varchange} shows that the divergent parts of $\Sigma_n$ and $\Phi_n$ are the $m=n$ terms, $\pi \lam T G_n$ and $\pi \lam T F_n$, respectively.  Since $G_n$ and $|F_n|$ are both of order 1 [see \eref{constraint}], these terms diverge in the strong coupling limit.
 
 All we have to do to avoid these divergences is to use \eref{Sb} instead of \eref{SB}. 
 Let us introduce
 reduced self-energies [cf. \eref{varchange}]
 \beg
 \begin{split}
\omega_n+ \Sigma'_n&=(\omega_n+\Sigma_n)-\pi T\lam G_n=\frac{b_n^z}{2\pi},\\
  \Phi'_n&=\Phi_n-\pi T\lam F_n=\frac{b_n^+}{2\pi}.
  \end{split}
  \label{bnz}
 \en
$\Sigma'_n$ and $\Phi'_n$ are the same as $\Sigma_n$ and $\Phi_n$ in \eref{varchange} but without the $n=m$ terms. It follows from \esref{Sb} and the definition~\re{bn} of $\bm b_n$ that
 \beg
\begin{split}
\Phi'_n= \pi  T \sum_{m\ne n}  \lambda_{nm}\frac{ \Phi'_m}{ \sqrt{(\omega_m+\Sigma'_{m} )^2 +| \Phi'_{m} |^2} },\\
\Sigma'_n= \pi  T \sum_{m\ne n}  \lambda_{nm}  \frac{\omega_m+ \Sigma'_m}{ \sqrt{(\omega_m+\Sigma'_{m} )^2 +| \Phi'_{m} |^2}},
\end{split}
\label{eliunfamiliar}
\en
where for simplicity we took $e_n=+1$ (no spin flips). As we discuss below, this is always the case at not too strong coupling and only   these $e_n=+1$ solutions   have been considered in the literature until now.  Then, \esref{eliunfamiliar} are equivalent to the original
Eliashberg equations~\re{elifamiliar3}, but do not contain $n=m$ terms. 
Note also that \esref{Sb} and \re{SB} imply $S_n^z/S_n^+=b_n^z/b_n^+=B_n^z/B_n^+$ and therefore
\beg
\frac{\Phi_n}{\omega_n+\Sigma_n}=\frac{\Phi'_n}{\omega_n+\Sigma'_n}.
\label{ppss}
\en

The $z$-component of spin, $S^z(\omega_n)\equiv S_n^z=G_n$, is odd in $\omega_n$, because $G_n$ is odd. Now consider an equilibrium configuration such that 
$\mbox{sgn}(S_n^z)=\mbox{sgn}(\omega_n)$. Taking into account that $S_n^z$ is odd, the expression~\re{bn} for $\bm b_n$ and
\eref{Bdef} imply  $\mbox{sgn}(b_n^z)=\mbox{sgn}(B_n^z)=\mbox{sgn}(\omega_n)$. Since  $b_n^z$ and $S_n^z$ have the same sign, this must be  a spin configuration  with no spin flips.   Though less obvious, the converse is also true, i.e., $\mbox{sgn}(b_n^z)=\mbox{sgn}(B_n^z)=\mbox{sgn}(\omega_n)$ holds in any configuration with no spin flips. 

 It is also not difficult to show that $\mbox{sgn}(b_n^z)=\mbox{sgn}(\omega_n)$ continues to hold when only a small number of spins are flipped out of this configuration (which we will assume). On the other hand, we saw above
that $\bm S_n$ must be parallel to $\bm B_n$ in any   spin equilibrium that corresponds to a solution of Eliashberg equations. Therefore, for such equilibria with few spin flips we have
\beg
\mbox{sgn}(B_n^z)=\mbox{sgn}(S_n^z)=e_n\mbox{sgn}(\omega_n).
\label{flip}
\en

\subsection{Spin-flip solutions}
\label{spin_flip_subsec}

Let us determine when spin-flip solutions of Eliashberg equations exist. The spin chain has equilibria with any number of spins at arbitrary positions being antiparallel to their fields. However, to be also solutions of the Eliashberg equations these configurations must  satisfy the inequality~\re{flipcond}. With the help of \eref{Sb} we rewrite this inequality as
\beg
e_n|\bm b_n|+ 2\pi^2\lam T\ge 0.
\label{ineq11}
\en
First, it is clear that $e_n=+1$ always works, i.e., stationary configurations of spins (equilibria) with no spin flips are always solutions of the Eliashberg equations. 

Now suppose $\bm S_n$ is antiparallel to its effective field $\bm b_n$. Then, $e_n=-1$ and \eref{ineq11} becomes
\beg
 2\pi^2\lam T\ge |\bm b_n|.
\label{ineq22}
\en
We see that no spin flips are allowed at weak coupling, $\lam\to0$, and at $T=0$ for any finite $\lam$, because $2\pi^2\lam T$ in this inequality   is either negligible or zero, while the right-hand side is positive. Conversely, arbitrary spin flips are permitted in the strong coupling limit $\lam\to\infty$ at any finite $T$, because this term   is infinite and positive. Therefore, we expect that at finite temperature, spin-flip solutions first appear at $\lam\sim1$.  

The existence of spin-flip solutions in the double limit $\lam\to\infty$ and $T\to 0$ depends on whether the  product $\lam T$ goes to zero (spin-flip solutions do not exist), infinity (solutions with any set of $\bm S_n$ antiparallel to $\bm b_n$ exist), or a finite value (spin-flip solutions may or may not exist). Recall that $\lam = g^2/\Omega^2$ and that we are increasing $\lam$ by decreasing $\Omega$. Then, arbitrary spin flips are allowed as long as $Tg/\Omega^2\to\infty$ in the double limit $T\to0$ and $\Omega\to 0$ and there are numerous such solutions in the regime $T\gg \Omega^2/g=g/\lam=\Omega/\sqrt{\lam}$. 

We mentioned in the previous subsection that  $m=n$  contributions to the self-energies $\Sigma_n$ and $\Phi_n$ in \esref{varchange}   diverge in the  strong coupling limit. It is precisely these divergent terms that make spin-flip solutions possible. Indeed, 
 \esref{Bdef}--\re{flipcond} and the definition  of $\bm b_n$ in \eref{bn} show that    the second term in the inequality~\re{ineq11}, without which the spin-flip solutions would not exist,  arises from just these two terms. 

Spin-flip solutions occur already in the normal state where $\bm S_n=\mbox{sgn}(\omega_n) \hat {\bm z}$. Suppose we flip the spin at Matsubara frequency $\omega_k>0$. Since $S_n^z$ is
odd in $\omega_n$, we also have to symmetrically flip the spin at $-\omega_k$. The new spin configuration is: $\bm S_n=\mbox{sgn}(\omega_n) \hat {\bm z}$ for $|\omega_n|\ne \omega_k$ and  $\bm S_n=-\mbox{sgn}(\omega_n) \hat {\bm z}$ for $|\omega_n|= \omega_k$.  Let us use the Holstein model~\re{schol} for
simplicity. The inequality~\re{ineq11} becomes
\beg
\begin{split}
\lam\ge 2k+1
+\sum_{m=1}^{k+1} \frac{2 g^2}{ 4\pi^2 T^2 m^2+\Omega^2}.
\end{split}
\label{lam_min}
\en
We see that the first spin-flip solution of the Eliashberg equations appears at $\lam\ge1$ for $k=0$ and $T\to\infty$, the next one at $\lam\ge 3$, etc. A similar analysis  in the normal  and   superconducting states at lower $T$    again shows that these solutions appear
at $\lam\gtrsim 1$ and proliferate at larger $\lam$.

Let us also determine the lowest temperature at which normal ($\Delta_n\equiv0$) solutions with spins antiparallel to their fields exist. It costs least to flip the spins at the first two Matsubara frequencies $\omega_{-1}$ and $\omega_0$. Setting $k=0$ in \eref{lam_min} and keeping in mind that $\lam=g^2/\Omega^2$, we find
\beg
T\ge \frac{\Omega}{2\pi}\sqrt{ \frac{\lam+1}{\lam-1} }.
\label{T_min_sf}
\en
At strong coupling, this becomes $T\ge\Omega/(2\pi)$, which implies the condition $T\gg \Omega/\sqrt{\lam}$ discussed above. Note also that in the strong coupling limit $\Omega\to0$, the quantity $\lam T$ diverges as $g^2/\Omega$ or stronger when $T$ satisfies the inequality~\re{T_min_sf}.
 
We argue  in Sec.~\ref{eli_kinetic} that spin-flip solutions  are important for understanding the short-time  nonlinear dynamics of strongly coupled superconductors: most of them are unstable fixed points of kinetic equations, similar to an inverted pendulum, that generate  soliton-like waves with a rich structure. 
 As mentioned in the Introduction, the Migdal-Eliashberg theory breaks down at a certain finite $\lam_c$.  Interestingly, both the emergence of  the spin-flip saddle points at  $\lam<\lam_c$ and the breakdown of the theory originate from the same source. We saw in this section that it is the divergent $n=m$ terms in the self energies that bring about the spin-flip solutions. In a subsequent paper~\cite{breakEli} we will see that these terms are also among the main culprits responsible for the breakdown of the Migdal-Eliashberg theory. In this connection,
 let us mention the mountain pass theorem~\cite{mountain_pass}  according to which there must be a saddle point between two minima. Therefore, it  may be
 that the emergence of these saddle points indicates that the free energy develops an additional minimum aside from the one described by the Eliashberg equations.

\section{Spins and superconducting gap function}
\label{gapsect}

 In this section, we introduce the superconducting gap function $\Delta_n$, write down a new, more general gap equation that accounts for spin flips, and   relate spin components to the gap function.

We introduce the Eliashberg gap function $\Delta(\omega_n)\equiv\Delta_n$ through equations
\beg
\omega_n+\Sigma_n =\omega_n Z_n,\quad  \Phi_n=\Delta_nZ_n. 
\label{subst}
\en
Equations~\re{bps} and \re{flip} imply
\beg
 \mbox{sgn}(Z_n)=e_n.
\en
In other words, $Z_n$ is negative if spin $\bm S_n$ is flipped and positive otherwise. As far as we are aware, only   solutions with  positive $Z_n$, i.e., with no spin flips have been considered in the literature until now.

Since $\Sigma_n$ is odd and $|\Phi_n|$ is even in frequency, $Z_n$ and $|\Delta_n|$ are both even.
With the substitution~\re{subst}, the Eliashberg equation for $\Sigma_n$ in~\re{elifamiliar3}  determines $Z_n$ for a given $\Delta_n$,
\beg
Z_n=1+\frac{\pi T}{\omega_n}\sum_m \lambda_{nm}\frac{e_m\omega_m}{\sqrt{\omega_m^2+|\Delta_m|^2}},
\label{Z}
\en
while the  equation for $\Phi_n$   becomes the gap equation
\beg
\omega_n\Delta_n=\pi T\sum_m \lambda_{nm}\frac{ \omega_n e_m\Delta_m - \Delta_n e_m\omega_m}{\sqrt{\omega_m^2+|\Delta_m|^2}},
\label{gapeq}
\en
where we used \eref{Z} to eliminate $Z_n$. As soon as the sign of $Z_n$ is fixed, the gap equation decouples from $Z_n$ and the Eliashberg equations reduce to a single equation: the gap equation~\re{gapeq}. The gap equation and the gap function determine various thermodynamic properties of the system, for example, the specific heat,  superconducting $T_c$, and the condensation energy. Continued to  real frequencies, the gap function determines the density of states and various response functions, such as the optical conductivity. 

The $m=n$ term vanishes in the gap equation~\re{gapeq}. This equation is therefore divergence free in the limit $\lam_{nn}=\lam\to\infty$. The same is not  true for $Z_n$ which diverges in this limit, because the $m=n$ term does not cancel from \eref{Z}. To get rid of this divergence, we use the reduced self-energies defined in Sec.~\ref{newsect} and introduce $Z_n'$ and $\Delta_n$ through
\beg
\omega_n+\Sigma'_n =\omega_n Z'_n,\quad  \Phi'_n=\Delta_nZ'_n. 
\label{subst'}
\en
Equation~\re{ppss} guarantees that $\Delta_n$ here is the same as in \eref{subst}. Therefore, the gap equation remains the same, while the expression for $Z'_n$ is
\beg
Z'_n=1+\frac{\pi T}{\omega_n}\sum_{m\ne n} \lambda_{nm}\frac{e_m\omega_m}{\sqrt{\omega_m^2+|\Delta_m|^2}}.
\label{Z'}
\en
The only difference between $Z'_n$ and $Z_n$ is that the $m=n$ term is absent in $Z'_n$.

Spin components in terms of the gap function are
\beg
S_n^z=\frac{ e_n \omega_n}{\sqrt{\omega_n^2+|\Delta_n|^2}},\quad S_n^+=\frac{ e_n \Delta_n}{\sqrt{\omega_n^2+|\Delta_n|^2}},
\label{spintogap}
\en
as follows  from \esref{fg} and \re{sgf}. Since the interaction is purely ferromagnetic, it is clear that for fixed $S_n^z$ the free energy is minimal when the $xy$  projections of all spins, $\bm S^\perp_n$, point in the same direction. Without loss of generality we take this direction to be the $x$ axis. The spin Hamiltonian~\re{spinh} becomes
\beg
H_s =-2\pi \sum_n \omega_n \cos\theta_n-\pi^2 T\sum_{nm} \lambda_{nm}\cos(\theta_n-\theta_m),
\label{htheta}
\en
where $\theta_n$ is the angle the spin $\bm S_n$ makes with the $z$ axis, $S_n^z=\cos\theta_n$ and $S_n^x=\sin\theta_n$. The  stationary point of  $H_s$ with respect to $\theta_n$ is
\beg
\omega_n\sin\theta_n= \pi T \sum_{m} \lambda_{nm}\sin(\theta_m-\theta_n).
\label{43}
\en
With the help of \eref{spintogap}, we see that this is nothing but the gap equation~\re{gapeq} written in terms of the angular variable $\theta_n$ and that $\Delta_n=\omega_n\tan\theta_n$. Since $|\sin(\theta_m-\theta_n)|\le 1$ and $\lam_{nm} \le g^2/(\omega_n-\omega_m)^2$ [see \eref{1morelam}], the right-hand side of \eref{43} is bounded in absolute value. It follows that $\sin\theta_n\to0$
when $\omega_n\to\pm\infty$. Feeding this information into the right-hand side, we see that moreover $\omega_n\sin\theta_n\to0$
and therefore $\Delta_n\to 0$ as $\omega_n\to\pm\infty$.

 Let us also mention here various expressions for the free energy of boson-mediated superconductors that have been proposed over the years.  Eliashberg derived one such expression   in 1962~\cite{elifree} (see also Ref.~\cite{wada}). Notably, Bardeen and Stephen used Eliashberg's result to evaluate the corrections to the BCS condensation energy due to finite   ratio of the BCS gap to the Debye energy~\cite{bardeen}. We discovered the spin-chain representation of the free energy~\re{schol}   in 2002 announcing it afterwards at
various venues~\cite{andrey_erratum, aps_abstract}. To reconcile our result with Eliashberg's,  Haslinger and  Chubukov obtained a somewhat less general (with all $e_n=+1$) answer for the free energy of the Holstein model in terms of $\Delta_n$  by integrating Eliashberg's expression over the electron momentum~\cite{andrey_f} . This answer follows from~\eref{schol}  if we substitute \eref{spintogap} with  $e_n=+1$ into it. Subsequently, this answer has been used in a number of publications. 

It is important to emphasize that our  expression~\re{feH1} for the free energy is more general than~\eref{schol}  and therefore Eliashberg's expression.  Equation~\re{feH1} takes into account certain fluctuations of the fields $\Sigma$ and $\Phi$ in imaginary time which \eref{schol} does not.   The two expressions are equivalent only at $\lam=\infty$, while away from this limit they coincide only at  the stationary points. We will discuss this in more detail below.

\section{Stationary points of the free energy}
\label{stptsec}

 This section contains  several results regarding stationary points of the free energy. 
One  is that at the minimum of the free energy $\Delta(\omega_n)=e^{i\phi}|\Delta_n|$,
where $|\Delta_n|$ is even in $\omega_n$ and $e^{i\phi}$ is an overall phase. In other words, up to an overall phase $\Delta(\omega_n)$ is non-negative and even. Another is that any spin configuration where $\Delta(\omega_n)$ is real, but changes sign, cannot be a
local or global minimum and is higher in energy than the global minimum by an amount proportional to the system size. In particular, spin flips correspond to saddle points. Fixing the overall phase, we also show that there cannot be a continuous family of stationary points connected to the global minimum and argue based on symmetry that the global minimum is unique.

We have seen in Sec.~\ref{newsect} that stationary points of the free energy are the equilibria of the spin chain. At weak coupling only
 equilibria where each spin is parallel to its magnetic field ($e_n=+1$) are stationary points. As $\lam$ increases, equilibrium configurations with arbitrary number of spins antiparallel to their fields ($e_n=-1$) become stationary points as well. These spin-flip solutions of the Eliashberg equations are necessarily saddle points. Indeed, suppose $\bm S_n$ is antiparallel 
to $\bm b_n$. Then, its contribution $h_n=- \bm b_n\cdot \bm S_n$ to the spin Hamiltonian is positive and decreases when we rotate
$\bm S_n$ keeping all other spins fixed.   Given that there are also spins parallel to their fields, the contribution of any of them increases when it deviates from its equilibrium position. Therefore, such solutions cannot be local minima or maxima, but are saddle points.

We see that at a minimum all spins must be parallel to their fields.
  This condition is only necessary, but not sufficient as there can be a collective mode involving many spins that lowers the free energy. This happens, for example, when the normal state loses stability at $T=T_c$. In the normal state all spins are   parallel to their fields and yet it is a saddle point and not a minimum below $T_c$.  Hence, for the minimum   we take $e_n=+1$ for all $n$ in \eref{spintogap}:
\beg
S_n^z=\frac{ \omega_n}{\sqrt{\omega_n^2+|\Delta_n|^2}},\quad S_n^+=\frac{ \Delta_n}{\sqrt{\omega_n^2+|\Delta_n|^2}},
\label{spintogap1}
\en
and  
\beg
\omega_n\Delta_n=\pi T\sum_m \lambda_{nm}\frac{ \omega_n\Delta_m - \Delta_n \omega_m}{\sqrt{\omega_m^2+|\Delta_m|^2}}
\label{gapeqmin}
\en
is the corresponding gap equation.

Before we proceed to analyze the global minimum, it is helpful to summarize the main features of the spin-chain representation of the free energy $f$:

\begin{enumerate}[label=(\alph*), left=0pt, itemsep=1pt, topsep=1pt]

\item The spin-chain formula, $f=\nu_0 T H_s$, for the free energy holds at stationary points of $f$. It does not necessarily hold away from such points because we used the Eliashberg equations~\re{elifamiliar3} in deriving it. However, the strong coupling limit, $\lam\to\infty$, is an exception. As we  show in Sec.~\ref{strongsect}, in this limit $f=\nu_0 T H_s$ holds at every point $(G_n, F_n)$ of the configuration space of the system,  not only at the stationary points.

\item Every stationary point of $f$ is also a stationary point of $H_s$ and every stationary point of $H_s$ with $e_n=+1$ is a stationary
point of $f$. Most importantly, the global minimum of $f$ is the global minimum of $H_s$. We also saw in Sec.~\ref{newsect} that in the $\lam\to\infty$ limit $e_n$ are unconstrained and, therefore, the correspondence between stationary points of $f$ and $H_s$ is one to one.

\end{enumerate}

 For easier reference, let us  write down the spin-chain Hamiltonian~\re{spinh21} here in a slightly different form
\beg
H_s =-2\pi \sum_n \omega_n S_n^z-\pi^2 T\sum_{nm} \lambda_{nm}({\bm S_n}\cdot {\bm S_m}-1),
\label{spinh211}
\en
and remind that $\lam_{nm}>0$ for all $n$ and $m$. In \eref{spinh211}  we regularized the otherwise divergent interaction term by subtracting a constant from  it~\cite{zeta}. This does not affect the stationary point (gap) equation  in any way.

\subsection{ Global minimum of the free energy}

 Here we establish the following properties of the free energy of the electron-phonon system:

\begin{enumerate}[label=(\arabic*), left=0pt, itemsep=1pt, topsep=1pt]

\item \label{prop1} At the global minimum $\Delta(\omega_n)$ is a   non-negative and even function of $\omega_n$ up to an overall phase. Moreover,
either $\Delta(\omega_n)=0$ for all $n$ (normal state) or  $\Delta(\omega_n)>0$ for all $n$ (superconducting). 

\item \label{prop2} For a fixed overall phase, there are no curves of stationary points  that contain the global minimum.  

\item \label{prop3} The only symmetry of the spin chain is the symmetry with respect to rotations around the $z$ axis.  

\item \label{prop4} Any $\Delta(\omega_n)$ that is real, but changes sign with $\omega_n$, cannot be a minimum (local or global) of $H_s$.

\end{enumerate}

These statements remain true at arbitrary temperature, including $T=0$~\cite{trueT0}.   Let us begin with the  proof of property~\ref{prop1}.
Suppose $S_n^z$ are fixed. Then, we have to minimize the $xy$  part of the spin Hamiltonian. Clearly, since $\lam_{nm} >0$ for all
$n$ and $m$, this requires all $\bm S^\perp_n$ being aligned in the same direction.  Any deviation from this alignment means a 
finite-energy cost in $H_s$ and because the total free energy   is $Nf=\nu_0 N T H_s$, the total-energy cost is proportional to the total number of sites $N$.  
As before  we designate the direction in which $\bm S^\perp_n$  align  to be the positive $x$ direction.  This sets the arbitrary overall phase  factor of $\Delta(\omega_n)$, $e^{i\phi}=1$.  Equation~\re{spintogap1} then implies that $\Delta_n$ is real and, most importantly, it must be  non-negative for all $n$ at the global minimum.  Since $|\Delta_n|$ is even, $\Delta_n=|\Delta_n|$ must also be an even function of frequency. Therefore, $\Delta(\omega_n)$ is a   non-negative and even function. Given that $\Delta_n$ is non-negative, the gap equation~\re{gapeqmin} further shows that either all $\Delta_n$ are zero or none.

 Clearly, if $\Delta_n$ is the global minimum, so is $e^{i\phi}\Delta_n$. Note that $e^{i\phi}\Delta_n$ defines a closed curve (circle) with no end points in the configuration space. Let us show that there can be  no other one-parameter families of solutions of the Eliashberg equations that contain the global minimum [property~\ref{prop2}]. Suppose such a family exists. It then similarly defines a curve in the configuration space and both $f$ and   $H_s$ must be constant along this curve. It follows that all their derivatives vanish on this curve as well. 
We saw above that at the global minimum we can always fix the global phase so that $S_n^y=0$ and $S_n^x\ge 0$.  Equation~\re{spintogap1} shows that $S_n^z\ge 0$ for $\omega_n\ge 0$. Therefore, the angle $\theta_n$ that spin $\bm S_n$ makes with the $z$ axis is in the range $[0, \pi/2]$. Then, the one-parameter family in question must have end points. As all derivatives of $H_s$ vanish along the curve of solutions before the end point and at least some of the first derivatives must be nonzero after the end point, $f$ cannot be an analytic function of all $\theta_n$.
This contradicts \eref{htheta}, which shows that $H_s$ is analytic in all $\theta_n$~\cite{subtract}. 

The model~\re{spinh211} is an inhomogeneous Heisenberg spin chain. Conditions under  which this type of models  are integrable for both quantum and classical spins have been  studied in the literature~\cite{gaudin,sklyanin0,sklyanin,bcs_dynamics,ortiz2005,gritsev}. In particular, there is a class of integrable models  
  that describe BCS-like pairing between fermions, which become spin chains when written in terms of Anderson spins~\cite{bcs_dynamics,ortiz2005,gritsev}. These models,
  integrable for  both quantum and classical spins, are  similar to the Hamiltonian~\re{spinh211} in that there is an inhomogeneous Zeeman field linear in the spin coordinate and interactions between spins are  ferromagnetic.   Models of this type are integrable only for   special choices of coupling
  constants $\lam_{nm}$. Integrable cases are a subset of measure zero among all Hamiltonians of this form, because we need only of order $N_s$ parameters to specify $\lam_{nm}$ in integrable cases  as opposed to $N_s^2$ in the generic case, where $N_s$ is the number of spins.
 It is straightforward to verify that our $\lam_{nm}$ do not correspond to any known integrable model. 
 It is  safe to conjecture that the Hamiltonian~\re{spinh211} is not integrable in the sense of Ref.~\cite{integrability}, i.e., that it does not possess parameter-dependent integrals of motion~(parameters here are $\omega_n$ and $\lam_{nm}$). Then,  any symmetry  must Poisson-commute  with each term in the Hamiltonian  individually, i.e., with each $S_n^z$ and $\bm S_n\cdot \bm S_m$. The only such symmetry is the
  $z$ component of the total spin $J_z=\sum_n S_n^z$. Thus, we arrive at  property~\ref{prop3} assuming non-integrability of the spin chain~\re{spinh211}. 
  
  Note that conservation of $J_z$ is responsible for the degeneracy of the global minimum with respect to rotations around the $z$ axis, i.e., with respect to the overall phase of $\Delta_n$. Since there are no other symmetries, we expect  the global minimum to be generally unique (barring accidental degeneracy) apart from the arbitrary overall phase of $\Delta_n$.

Now consider a spin configuration with real sign-alternating $\Delta(\omega_n)$. We already know from property~\ref{prop1} that such a configuration cannot be a global minimum and is associated with a macroscopic energy cost.  Let us investigate if it can at least be a local minimum. Equation~\re{spintogap1}  implies that the projection of spin $\bm S_n$ onto the $xy$  plane is $\bm S_n^\perp=\mbox{sgn}(\Delta_n)|\bm S_n^\perp|\hat {\bm x}$, where $\hat {\bm x}$ is the unit vector along the $x$ axis. Let us uniformly rotate all spins with $\mbox{sgn}(\Delta_n)<0$ around the $z$ axis towards the positive $x$ axis by a small angle
$\delta\phi$, i.e., for these spins $S_n^+\to S_n^+ e^{i\delta\phi}$. This does not change the Zeeman energy and the interaction energies of
spins with $\mbox{sgn}(\Delta_n)<0$ among themselves and of spins with $\mbox{sgn}(\Delta_n)\ge0$ among themselves, while the
ferromagnetic interaction energy of $\mbox{sgn}(\Delta_n)<0$ spins   with $\mbox{sgn}(\Delta_n)>0$ decreases
as they become more aligned with each other. Since the energy decreases for an infinitesimal deviation from this configuration, it cannot be a local minimum of $H_s$. At $\lam=\infty$, when  $f=\nu_0 T H_s$ holds in the entire configuration space, spin configurations with  sign-alternating $\Delta(\omega_n)$ cannot be local minima of $f$ as well. This proves property~\ref{prop4}.

In connection with the above, let us mention  papers by Wu \textit{et. al.} that claim that  at $T=0$ and $\lam=\infty$ there is a 
one-parameter family of sign-alternating solutions of the Eliashberg equations~\cite{andreygamma2,andrey_gamma23,note234}. One end point of this  purported family of solutions is said to be the global minimum and the other, the normal state, i.e., the free energy must \textit{increase} along this   continuous family of stationary points (see, e.g., Fig.~4 of Ref.~\cite{andreygamma2} and Fig.~1 of Ref.~\cite{andrey_gamma23}). This is impossible because any curve of stationary points must also be a curve of constant free energy, such as, e.g., the curve traced out by changing the global phase $\phi$. It also contradicts property~ \ref{prop2} above.   
 
Moreover, we saw above that the energy cost of configurations with sign-alternating $\Delta_n$ is proportional to the system volume. 
 Their Boltzmann weight relative to the global minimum is $e^{-N\Delta f/T}$, where $N$ is the number of sites and $\Delta f$ is the free-energy density difference (which is finite in $N\to\infty$ limit).  Naturally, such states cannot contribute to thermal equilibrium properties of a bulk system. In the present case not only $N\to\infty$, but also $T\to0$, so the global minimum becomes the ground state and entirely determines all equilibrium physics.

\section{Strong coupling limit}
\label{strongsect}

There are two universal limits of the Migdal-Eliashberg theory: the weak and strong coupling limits, $\lam\to0$ and $\lam\to\infty$. ``Universal''  here means that the theory becomes independent of the phonon dispersion and momentum dependence of the electron-phonon coupling. In both limits there is only one low-energy scale. The weak coupling limit of the Migdal-Eliashberg theory is the BCS theory (see Appendix~\ref{weak}). 

 We review the strong coupling limit~\cite{allendynes,carbotte,strong3,strong1,strong2,combescot} in this section for two related reasons. First, the derivation of the spin-chain representation of the free energy does not 
rely on  Eliashberg equations in this limit and therefore holds everywhere in the configuration space.   Second, we will need  it in a later study. However, it is important to emphasize that the strong coupling limit of the Migdal-Eliashberg theory is \textit{unphysical}, because the theory breaks down at finite $\lam$ before reaching $\lam=\infty$ as discussed in the  Introduction.
  We will use the strong coupling limit  as a technical tool only, e.g.,  to show that the breakdown occurs universally, irrespective of the model electron-phonon Hamiltonian.

Consider the Holstein model first. By definition
\beg
 \lambda=\frac{g^2}{\Omega^2}=\frac{\nu_0 \alpha^2 M^{-1}}{\Omega^2},
 \en
where $\alpha$ is the dimensionful electron-phonon coupling constant. Note that $\alpha$ does not have units of energy, while $g$ does. We define the strong coupling limit as $\alpha\to\infty$. This implies $\lam\to\infty$ and $\alpha, g\propto \lam^{\frac{1}{2} }$. As $\Omega$ is negligible  compared to $g$, the latter remains the only intrinsic low-energy scale in the problem~\cite{high}. Other energies characterizing the superconductor are proportional to $g$, for example, $T_c\approx 0.18g$ and the spectral gap is $1.16g$~\cite{allendynes,combescot}. Measured in units of $g$ these quantities are finite, while the phonon frequency goes to zero, $\Omega\propto\lam^{-\frac{1}{2} }$. Therefore, an equivalent, but more convenient  way to obtain the strong coupling limit is to keep $g$ fixed and send $\Omega$ to zero.

We already saw that   $Z_n$ and therefore $\Phi_n$ and $\Sigma_n$ diverge when $\lam\to\infty$, while $F_n$ and $G_n$ stay finite. 
Separating the divergent part in the variable change~\re{varchange}, $\Sigma_n=\lam\pi T G_n +\dots$,  $\Phi_n=\lam\pi T F_n +\dots$
and expanding the square root in \eref{feH1} in $\Omega^2$, we directly obtain the spin Hamiltonian~\re{spinh} up to terms of order $\Omega^2$ that vanish in the strong coupling limit. Moreover, it turns out that the mass of fluctuations violating  the constraint~\re{constraint} is infinite~\cite{meaningmigdal} and hence $\bm S_n^2=1$. Note that here we cast the free energy into the spin chain form without ever using the stationary point (Eliashberg) equations. Therefore, in the strong coupling limit the spin-chain representation is guaranteed at any point $(G_n, F_n)$ in the configuration space with no stationary point constraints on  $G_n$ and $F_n$.
The spin-chain Hamiltonian~\re{schol} in the strong coupling limit, $\Omega\to0$, becomes
\beg
H_s=-2\pi \sum_n \omega_n S_n^z-\pi^2 Tg^2\sum_{nm} \frac{ {\bm S_n}\cdot {\bm S_m}-1}{(\omega_n-\omega_m)^2},
\label{schol1}
\en
where we subtracted a constant from the Hamiltonian as discussed below \eref{spinh21}.

The  procedure for dispersing phonons is the same. Now
\beg
\begin{split}
 \lambda(\omega_l)&=\frac{1}{2p_F^2} \int_0^{2p_F} \!\!\!\!  \frac{ g_{q}^2 q dq}{\omega_l^2+\omega_{q}^2},\\
  \lambda&=\lambda(\omega_{l}=0)=\frac{1}{2p_F^2} \int_0^{2p_F}   \frac{  g_{q}^2 q dq}{ \omega_{q}^2},
  \end{split}
 \en
 see Appendix~\ref{standardapp}.
 Suppose $g_q\to\infty$ for some range of $q$. Then, the energy $g$ defined as
 \beg
 g^2\equiv \frac{1}{2p_F^2} \int_0^{2p_F} \!\!\!\!      g_{q}^2 q dq
 \en
diverges and all other low energies are either negligible or  proportional to it. This is equivalent to keeping $g$ fixed and sending the phonon frequencies $\omega_q$ to zero. In this limit, the spin-chain Hamiltonian~\re{spinh} for dispersing phonons  turns into the Hamiltonian~\re{schol1}, same as for the Holstein model. We see that the spin chain~\re{schol1}  provides a complete and universal description of the the thermodynamics of the strong coupling limit: it determines the Boltzmann weight of any field configuration and  is   independent of the underlying electron-phonon model except for a single energy $g$.

\section{Comparison to  Anderson pseudospins} 
\label{ander}

We already mentioned in Sec.~\ref{spinsect} that the description of the normal and superconducting states and of the transition between them in terms of spins introduced in this paper is   very similar to the Anderson pseudospin description of the BCS theory of  superconductivity~\cite{pseudo}. In particular, our Fig.~\ref{spinsfig} is   identical to Fig.~1 of Ref.~\cite{pseudo}, except that in the latter figure the sites of the chain are single-fermion energies $\xi_\pp$ instead of fermionic Matsubara frequencies $\omega_n$.  Indeed, we will see in this section that equilibrium configurations of Eliashberg and Anderson spins map into each other under the interchange of $\xi_\pp$ with $\omega_n$ and of BCS gap $\Delta$ with the Eliashberg gap function $\Delta_n$.

An overall principle common to both spin formulations of superconductivity is that there are three real functions of energy or frequency (the normal average and  real and imaginary parts of the anomalous average) that admit an interpretation  as three components of a classical spin. Nevertheless, Eliashberg and Anderson spins are distinct.  Eliashberg spins are energy-integrated normal and anomalous thermal Green's functions. They exist at any temperature and coupling  $\lam$. Anderson spins are frequency-integrated normal and anomalous Keldysh Green's functions (see below).  They are well-defined both in and out of equilibrium, but only at $T = 0$. The two sets of spins  
  do not coincide even where their domains of definition overlap, i.e.,  at $T=0$ in the 
weak coupling (BCS) limit of the Migdal-Eliashberg theory. 

\subsection{In and out of equilibrium BCS superconductivity in terms of classical Anderson pseudospins -- brief review}

Soon after the publication of the BCS theory, Anderson realized~\cite{pseudo} that the BCS Hamiltonian
\beg
H=\sum_{\pp\sigma} \xi_\pp c^\dagger_{\pp\sigma} c_{\pp\sigma}-\lam\delta \sum_{\pp\pp'} c^\dagger_{-\pp\up} c^\dagger_{\pp\dn} c_{\pp'\dn} c_{-\pp'\up},
\label{bcs}
\en
where $\delta=(\nu_0 N)^{-1}$, maps to a spin-$\frac{1}{2}$ model. We will need quantum averages of pseudospin-$\frac{1}{2}$  operators defined by Anderson
\beg
s_\pp^z=\frac{\langle c^\dagger_{\pp\up} c_{\pp\up} + c^\dagger_{-\pp\dn} c_{-\pp\dn}\rangle -1}{2},\quad s_\pp^-=\langle c_{\pp\dn} c_{-\pp\up}\rangle, 
\label{aspins}
\en
where $s_\pp^\pm=s_\pp^x \pm i s_\pp^y$. Note that $\bm s_\pp$  depend on time out of equilibrium, when the state of the system in which we evaluate the averages~\re{aspins} is itself time-dependent.

In the mean field treatment, which is exact for the BCS model in the thermodynamic limit~\cite{richardson,roman,baytin}, the Hamiltonian~\re{bcs} becomes~\cite{foster}
\beg
H=\sum_{\pp} 2\xi_\pp s_\pp^z - \lam\delta \sum_{\pp\pp'} s_\pp^+ s_{\pp'}^-.
\label{bcss}
\en
The eigenstates of the BCS Hamiltonian are product states of the form $\prod_\pp (u_\pp +v_\pp c^\dagger_{\pp\up} c^\dagger_{-\pp\dn}) |0\rangle$. For any state of this form
\beg
s_\pp^z=\frac{|v_\pp|^2-|u_\pp|^2}{2},\quad s_\pp^-=u_\pp^* v_\pp.
\label{uv}
\en
Therefore, the normalization condition $|u_\pp|^2+|v_\pp|^2=1$ implies that  $\bm s_\pp=(s_\pp^x, s_\pp^y, s_\pp^z)$ is a vector of length $\frac{1}{2}$.  Given the spin configuration, we determine the state of the system using \eref{uv}. 

Equilibria of the spin Hamiltonian~\re{bcss} correspond to the eigenstates of the BCS Hamiltonian~\cite{pseudo,solitons,osc,morep}. Spin $\bm s_\pp$ experiences an
effective magnetic field $\bm B_\pp=(-2\Delta_x, -2\Delta_y, 2\xi_\pp)$, where $\Delta_x$ and $-\Delta_y$ are the real and imaginary parts
of the BCS order parameter defined by
\beg
\Delta=\lam\delta \sum_\pp s_{\pp}^-.
\label{BCSgap}
\en
In equilibrium the spin is either antiparallel ($e_\pp=+1$) or parallel ($e_\pp=-1$) to its effective field. It follows that equilibrium spin configurations are
\beg
2s_\pp^z = -\frac{ e_\pp \xi_\pp}{\sqrt{\smash{\xi_\pp^2}+|\Delta|^2}},\quad 2s_\pp^- = \frac{ e_\pp \Delta}{\sqrt{\smash{\xi_\pp^2}+|\Delta|^2}}.
\label{bcseq}
\en
The self-consistency condition~\re{BCSgap} becomes
\beg
2\Delta= \lam\delta \sum_\pp \frac{ e_\pp \Delta}{\sqrt{\smash{\xi_\pp^2}+|\Delta|^2}}.
\label{bcsgapeq}
\en
The BCS ground state  is obtained by aligning all spins antiparallel to their fields, i.e., $e_\pp=+1$ for all $\pp$. States where one of
the spins is flipped opposite to its ground state orientation  correspond to excited or ``real'' pairs in the terminology of Bardeen, Cooper, and Schrieffer~\cite{bcs,pseudo}. Note that \esref{bcseq} and \re{bcsgapeq} describe two types of equilibrium configurations -- configurations with $\Delta=0$ [which is always a solution of  \eref{bcsgapeq}] and configurations  where $\Delta\ne0$. In the terminology of Ref.~\cite{osc}, these are the \textit{normal}    and  \textit{anomalous} states, respectively.

Classical Anderson pseudospins  play a central role in  understanding   collisionless  dynamics of BCS condensates in response to fast perturbations~\cite{foster,osc,solitons,turbulence,levitov,dzero,barankov,aoki}  when the sample size is of the order of the superconducting coherence length or smaller~\cite{turbulence}.
 Hamilton's equations of motion for the classical spin Hamiltonian~\re{bcss} with the usual angular momentum Poisson brackets for components of $\bm s_\pp$ are equivalent to the time-dependent Bogoliubov--de Gennes equations~\cite{solitons}. These classical spin equations of motion have been used by Anderson to analyze the collective modes of a BCS superconductor and more recently to study its time evolution after a quantum quench~\cite{foster,osc,levitov,dzero,barankov,aoki}.  Normal and anomalous states mentioned in the preceding paragraph play 
a special role in the   dynamics~\cite{solitons}. Normal states are   dynamically unstable and anomalous states are   unstable when sufficiently many spins are flipped. These unstable equilibria give rise to normal and anomalous multi-solitons that start in an unstable equilibrium at $t=-\infty$ and return into it at $t=+\infty$, similar to an inverted pendulum. Moreover,  dynamics for many physical initial conditions (e.g., for an interaction quench) can be described in terms of multi-solitons.

 Let us also mention the relationship between Anderson pseudospins and  equal time, zero-temperature  Keldysh Green's functions defined by~\cite{keldysh0,keldysh1,keldysh2}
\beg
\begin{split}
\mathsf{G}_\pp(t, t')=-i\langle [c_{\pp\sigma}(t), c_{\pp\sigma}^\dagger(t')]\rangle,\\
 \mathsf{F}_\pp(t, t')= -i\langle  [c_{\pp\up}(t), c_{-\pp\dn}(t')] \rangle,
 \end{split}
\en
where the square brackets denote the commutator and we assumed time-reversal symmetry, so that $\mathsf{G}_\pp(t, t')$ does not depend on $\sigma$. We see that $G_\pp(t,t)=2i s_\pp^z$ and $F_\pp(t,t)=2i s_\pp^-$. Fourier transforming the Green's functions with respect to $t'-t$, we find
\beg
2i s_\pp^z=\int_{-\infty}^\infty \!\!\!\!d\omega \mathsf{G}_\pp(\omega),\quad 2i s_\pp^-=\int_{-\infty}^\infty \!\!\!\!d\omega \mathsf{F}_\pp(\omega).
\label{angreen}
\en
We suppressed the  dependence on $t$ in spin components, $\mathsf{G}_\pp(\omega)$, and $\mathsf{F}_\pp(\omega)$.  Thus, we see that Anderson spins are frequency-integrated normal and anomalous Keldysh Green's functions (divided by $2i$).

\subsection{ Strongly coupled superconductors in and out of equilibrium in terms of Eliashberg spins}
\label{eli_kinetic}

Let us first compare equilibrium configurations of Eliashberg and Anderson spins. In equilibrium, Eliashberg spins are [\eref{spintogap}] 
\beg
S_n^z=\frac{ e_n \omega_n}{\sqrt{\omega_n^2+|\Delta_n|^2}},\quad S_n^+=\frac{ e_n \Delta_n}{\sqrt{\omega_n^2+|\Delta_n|^2}}.
\label{spintogap22}
\en
We see that they turn into Anderson spins~\re{bcseq}, up to a factor of 2 and opposite sign of $y$ and $z$ components, if we replace the Matsubara frequency $\omega_n$ with the single-particle energy $\xi_\pp$ and the Eliashberg gap function $\Delta_n$ with the BCS gap $\Delta$. The factor of 2 difference comes from  normalization, the arbitrary choice of spin length $\frac{1}{2}$ for Anderson spins and $1$ for Eliashberg spins. 
  Opposite signs of $y$ and $z$ components of spins are a similar ``gauge'' degree of freedom that does not affect the Poisson brackets between spin components. It is straightforward to redefine Anderson or Eliashberg spins to eliminate these differences. However,   the gap equations~\re{bcsgapeq} and \re{gapeq} are rather different due to the retarded nature of interactions and the presence of
   $zz$ interactions in the spin formulation of the  Migdal-Eliashberg theory.
  
 There are also similarities between the two approaches in the overall physical picture they provide. For example, in both cases the normal state is a state where all spins are down (up) below a certain energy and up (down) above it as shown in Fig.~\ref{spinsfig} and spins acquire $xy$  components  in the superconducting state thus softening the domain wall at zero  frequency (energy).  
 
 Nevertheless, the two sets of spins never coincide. Anderson spins are ill-defined away from $\lam\to0$, $T=0$ limit, i.e., in the Migdal-Eliashberg theory at finite $\lambda$ or  finite $T$, in the sense that the model cannot be formulated entirely in terms of them
 and the length of the vector $\bm s_\pp$ depends on $\pp$ and is not conserved by the dynamics. In contrast,  Eliashberg spins remain well-defined in the BCS ($\lam\to0$) limit and, in particular, the classical spin Hamiltonian~\re{spinh211} minus its value in the
normal state gives the BCS condensation energy as a function of $T$. Even in the BCS limit the  two sets of spins are distinct and provide alternative descriptions of the system -- Eliashberg spins determine the free energy as a function of $T$, while Anderson spins work at $T=0$ and determine the ground state and excited states of the Hamiltonian.

While Anderson spins  are frequency-integrated Keldysh Green's functions [see \eref{angreen}],  Eliashberg spins correspond to energy-integrated normal and anomalous Matsubara Green's functions
\beg
\begin{split}
\GG_{\sigma\pp} (\tau-\tau')=- \langle T_\tau c_{\pp\sigma}(\tau)  c^\dagger_{\pp\sigma}(\tau')\rangle,\\
\FF_\pp (\tau-\tau')= \langle T_\tau c_{-\pp\dn}(\tau)  c_{\pp\up}(\tau')\rangle.
\end{split}
\en
We have evaluated these Green's functions in the Matsubara frequency domain in Appendix~\ref{gfapp}:
\beg
\begin{split}
  \GG_{\pp n}=-\frac{ i(\omega_n+\Sigma_{n} )+ \xi_\pp +\chi_{n} } {(\omega_n+\Sigma_{n} )^2 +| \Phi_{n} |^2 + (\chi_{n} +\xi_\pp)^2  },\\
\FF_{\pp n}=-\frac{ \Phi_{n}   } { (\omega_n+\Sigma_{n} )^2 +| \Phi_{n} |^2 + (\chi_{n} +\xi_\pp)^2}.
\end{split}
\label{gfsym1}
\en
Integrating $ \GG_{\pp n}$ and $\FF_{\pp n}$ and comparing with the expressions~\re{fg} for Eliashberg spins on the stationary point, we see that
\beg
S_{n}^z=-i\int_{-\infty}^\infty\!\!\!\! d\xi_\pp \GG_{\pp n},\quad S_{n}^+=-\int_{-\infty}^\infty\!\!\!\! d\xi_\pp \FF_{\pp n}.
\label{eligreen}
\en
Note that the single-particle energy $\xi_\pp$ plays the same role for Eliashberg spins as the frequency $\omega$ does for Anderson spins.

 We explained in the previous subsection  how Anderson spins describe the far from equilibrium dynamics of a BCS condensate. Likewise,  we  expect  the time evolution of  Eliashberg spins with the Hamiltonian~\re{spinh211} to describe the dynamics of electronic degrees of freedom of a strongly coupled (Eliashberg) superconductor in a similar regime. Hamiltonian (Bloch) equations of motion for Eliashberg spins read as
 \beg
 \begin{split}
 \frac{d \bm S_n}{dt}&= -\bm b_n \times \bm S_n,\\
  \bm b_n&=2\pi\omega_n\hat {\bm z}+2\pi^2 T\sum_{m\ne n} \lam_{nm} \bm S_m.
  \end{split}
  \label{elibloch}
 \en
These equations or a certain version of them  should be valid  for the short-time far from equilibrium dynamics, at least when the sample size is of the order of the coherence length or smaller. In this case, dynamics are spatially uniform~\cite{turbulence},  phonons are in thermal equilibrium with the outside environment~\cite{eli_dynamics}, and collision integrals are negligible~\cite{vk}. Indeed, kinetic equations very similar to \eref{elibloch} have already been derived by Eliashberg~\cite{eli_dynamics} (see also Ref.~\cite{vk}). In the weak coupling limit, $\lam_{nm}\to\lam$ at relevant frequencies and,
with the replacement of $\omega_n$ with $\xi_\pp$, \eref{elibloch} is equivalent to the Bloch equation for  Anderson spins $\bm s_\pp$
defined by \eref{aspins}.

The spin-flip solutions we identified in Sec.~\ref{spin_flip_subsec} play a special role in the dynamics. These saddle points of free energy are by construction stationary solutions of the Bloch equation~\re{elibloch}. Just as with flipped Anderson spins discussed below \eref{bcsgapeq},
these equilibria are dynamically unstable when sufficiently many spins are flipped and there are two types of  them,  normal and anomalous, depending on whether or not the $xy$ components of spins are zero. Therefore, we anticipate that these saddle points should give rise to dynamics analogous to normal and anomalous solitons in the far from equilibrium BCS superconductivity.

\section{Conclusion}

In this paper, we mapped the Migdal-Eliashberg theory of superconductivity to a classical Heisenberg spin chain in Zeeman magnetic field. Lattice sites are fermionic Matsubara frequencies $\omega_n$. The interaction between spins is purely ferromagnetic. It depends on the phonon spectrum and falls off as $(\omega_n-\omega_m)^{-2}$ at large separation between spins. The Zeeman field is proportional to $\omega_n$, the spin coordinate along the chain. As an example, the spin Hamiltonian for the Holstein model is
\beg
H_s=-2\pi \sum_n \omega_n S_n^z-\pi^2 Tg^2\sum_{nm} \frac{ {\bm S_n}\cdot {\bm S_m}-1}{(\omega_n-\omega_m)^2+\Omega^2},
\label{scholconcl}
\en
where $T$ is the temperature, $g$ is the electron-phonon coupling, $\Omega$ is the renormalized Einstein phonon frequency, and  
$\bm S_n^2=1$. The free-energy density of the system of fermions interacting through phonons is $f=\nu_0 T H_s$.

The spin-chain formulation made the analysis of the free-energy functional  simple. In particular, we saw that, up to an overall phase $e^{i\phi}$, the   Eliashberg  order parameter $\Delta(\omega_n)$  is non-negative and even at the global minimum  and there can be no continuous families of stationary points that include the global minimum.  We also discussed symmetry and non-integrability arguments that further support our claim that the global minimum is unique.  

It became apparent that  the free energy acquires  infinitely many new saddle points (which are new, additional solutions of the Eliashberg equations) at strong coupling, which correspond to spin flips. We saw that these saddle points most likely play a significant role in the collisionless far from equilibrium dynamics of strongly coupled superconductors. Apparently, they  are the fixed points of the corresponding kinetic equations and give rise to rich multi-soliton dynamics.  The new saddle points emerge and proliferate   just before the breakdown of the Migdal-Eliashberg theory and, moreover, these two phenomena have similar origin as we discussed in Sec.~\ref{spin_flip_subsec}.
It is evident that the spin chain has  unstable equilibria where a number of spins are flipped against their magnetic fields. However, without the spin-chain representation these solutions are hard to notice and indeed they have not been seen before. 

Our spin-flip solutions of  Eliashberg equations   are   new and unrelated to the continuous family of sign-alternating solutions for $\Delta_n$ conjectured by Wu \textit{et. al.}~\cite{andreygamma2} at $\lam=\infty$. We  believe  that  this conjecture is internally inconsistent in several ways. Therefore, the claims by the aforementioned reference of vanishing of the superconducting $T_c$, destruction of the superconducting phase coherence, etc., are unsubstantiated.   
 Thinking in terms of the spin chain also made it clear that any sign-alternating $\Delta_n$  costs a macroscopic amount of energy to create
at all $\lambda$ and therefore cannot contribute to thermal equilibrium properties of a macroscopic system.  Most importantly, because as we discussed above the Migdal-Eliashberg theory breaks down at finite $\lam$, its  strong coupling limit is unphysical altogether.

Classical spins in the Migdal-Eliashberg theory (Eliashberg spins) are in many ways similar to    Anderson spins in the BCS theory of superconductivity, though the two sets of spins never coincide. Common to both spin  notions is that three real numbers, the normal Green's function and real and imaginary parts of the anomalous Green's function integrated over single-fermion energy (Eliashberg) or  frequency (Anderson), become the three components of a classical spin vector. Eliashberg spins exist at any temperature and any $\lam$, while Anderson spins are well defined in and out of equilibrium, but only at $T=0$ and in the BCS limit of the Migdal-Eliashberg theory. Equilibrium configurations of Anderson and Eliashberg spins map into each other under the interchange 
of single-fermion energy $\xi_\pp$ with fermionic Matsubara frequency $\omega_n$ and of the BCS gap $\Delta$ with Eliashberg gap function $\Delta_n$.
An interesting open question is to study the short-time far from equilibrium dynamics of strongly coupled conventional superconductors with the help of Eliashberg spins.

\vspace{5mm}
 
\begin{acknowledgments}

We thank Ar. Abanov, A. V. Chubukov, and M. K.-H. Kiessling for discussions and comments.

\end{acknowledgments}


\onecolumngrid 

\appendix 
\section{Eliashberg free energy for the Holstein model with arbitrary single-particle potential}
\label{holsteinapp}

In this appendix,  we derive the Eliashberg free energy  from the path integral for the Holstein model (see also Refs.~\cite{schmalian,onemorepath}). For the most part the derivation follows a standard sequence of steps. One notable difference is that  we keep the single-electron potential arbitrary throughout the entire calculation and rewrite the action in its eigenbasis. This will be especially useful later, in our study of the breakdown of the Eliashberg theory at finite $\lam$~\cite{breakEli}. 

\subsection{Effective action}

The  Lagrangian corresponding to the Holstein Hamiltonian~\re{holsteinH}   is
\beg
\begin{split}
L=  \sum_{\bm i \bm j, \sigma}  \psi^*_{\bm i\sigma} G_{0\ii\jj}^{-1} \psi_{\bm j \sigma}+
 \sum_{\bm i}\left[ \frac{M\Omega_0^2 \varphi^2_{\bm i} }{2} +
\frac{M (\partial_\tau \varphi_{\bm i} )^2}{2  } \right]
+
 \alpha \sum_{\bm i \sigma} \psi^*_{\bm i\sigma} \psi_{\bm i \sigma}\varphi_{\bm i}+\sum_{\ii\sigma} (J^*_{\ii\sigma}\psi_{\ii\sigma}+  \psi^*_{\ii\sigma} J_{\ii\sigma}),
\end{split}
\label{holaction}
\en
where $G_{0\ii\jj}^{-1}=\partial_\tau \delta_{\ii\jj}+h_{\bm i\bm j}-\mu\delta_{\ii\jj}$,  $ \psi_{\bm i \sigma}$ and $ \psi_{\bm i \sigma}^*$ are Grassmann variables, and $\varphi_{\bm i}$ is a real field that corresponds to the ion displacement operator $x_\ii$. We also introduced a chemical potential $\mu$ and Grassmann source fields $J^*_{\ii\sigma}$ and $J_{\ii\sigma}$, which we will use
to evaluate Green's functions. All fields in the Lagrangian depend on the imaginary time $\tau$. 

Integrating out phonons, we obtain the following effective action:
\beg
\begin{split}
S_\mathrm{eff}  =  \sum_{\ii\jj\tau\sigma} \psi^*_{\ii\tau \sigma } G_{0\ii\jj}^{-1} \psi_{\jj\tau \sigma } 
-\frac{1}{2}  \sum_{\bm i\tau\tau'\sigma\sigma'}  \psi^*_{\bm i\tau' \sigma' } \psi_{\bm i\tau' \sigma' } D_{\tau'\tau} \psi^*_{\bm i\tau \sigma } \psi_{\bm i\tau \sigma } 
  +\sum_{\ii\tau \sigma} \left( J^*_{\ii\tau \sigma  }\psi_{\ii\tau \sigma }+ \psi^*_{\ii\tau \sigma } J_{\ii\tau \sigma } \right)\!.
\end{split}
\label{S_fermion}
\en
 Summations  indicate summation over $\ii$ and integration over $\tau$ and $\tau'$,
 \beg
 D_{\tau'\tau}=T\sum_l \frac{  \alpha^2 M^{-1} }{\omega_l^2+\Omega^2} e^{i\omega_l (\tau'-\tau)},
 \label{Ddefinition}
 \en
 is the  effective electron-electron interaction, and  $\omega_l=2\pi T l$ are bosonic Matsubara frequencies. The interaction is proportional to   the phonon propagator as usual~\cite{agd}. 
 
 We replaced the bare phonon frequency $\Omega_0$ with the renormalized frequency $\Omega$. To obtain an equation for $\Omega$  within the path integral framework, one needs to introduce
 an additional  Hubbard-Stratonovich field $\Pi_\ii(\tau', \tau)$ for phonons~\cite{schmalian}. On the stationary point this  leads to the usual    phonon renormalization procedure (see Ref.~\cite{agd}, p. 181). We  reject it here, because Holstein and other standard electron-phonon Hamiltonians do not renormalize phonons correctly~\cite{kagan,geilikman}. Instead, we treat $\Omega$ as a parameter of the model. Renormalized phonon  frequencies are generally  momentum dependent even when the bare spectrum is dispersionless.  Then, the effective action is that for dispersing phonons, which we treat in Appendix~\ref{standardapp}. Here we   disregard the momentum dependence of $\Omega$ for simplicity: \eref{S_fermion} is a legitimate model of the electron-phonon system in its own right, no less legitimate than the Holstein model itself.

 We decouple the four-fermion term in \eref{S_fermion} with a Hubbard-Stratonovich transformation in the particle-particle and particle-hole channels with three fields $\Phi_\ii(\tau',\tau)$, $\Sigma_{\ii\up}(\tau',\tau)$, and $\Sigma_{\ii\dn}(\tau',\tau)$:
 \beg
\begin{split}
 S_\mathrm{eff}  = 
\sum_\ii \int\!\!\!\!\int d\tau'd\tau \left[ \frac{ \Phi^*_\ii (\tau',\tau) \Phi_\ii(\tau',\tau)}{ D(\tau'-\tau) } +\sum_{\sigma}\frac{ \Sigma_{\ii\sigma}(\tau',\tau) \Sigma_{\ii\sigma}(\tau,\tau')}{2D(\tau'-\tau)}  \right] +\\
 +\sum _{x'x} \Psi^\dagger_{x'}   M_{x'x}  \Psi_{x} 
  +\sum_{x}\left( K_{x}^\dagger \Psi_{x}+
\Psi_{x}^\dagger K_{x} \right)\!,
\end{split}
\label{S_Nambu}
\en
where $\Psi_x$ and $K_x$ are two-component Nambu fields, $M_{x'x}$ is a $2\times 2$ matrix in the Nambu space,
\beg
\Psi_x=\begin{bmatrix}
\psi_{\up x}\\
 \psi^*_{\dn x}\\
\end{bmatrix}\!,
\quad
K_x=
\begin{bmatrix}
J_{\ii\up}(\tau)\\
-J_{\ii\dn}^*(\tau)\\
\end{bmatrix}\!,
\quad
M_{x'x} =
\begin{bmatrix}
G_{0\ii\jj}^{-1}\delta(\tau'-\tau)-  i  \Sigma_{\ii \up}(\tau', \tau) \delta_{\ii \jj} &  \Phi_{\ii}(\tau', \tau) \delta_{\ii \jj}\\
\Phi_{\ii}^*(\tau, \tau')\delta_{\ii \jj}  & {\overbar{G}}_{0\ii\jj}^{-1}\delta(\tau'-\tau)+  i  \Sigma_{\ii\dn}(\tau, \tau') \delta_{\ii \jj}\\
\end{bmatrix}\!,
\en
and ${\overbar{G}}_{0\ii\jj}^{-1}=\partial_\tau \delta_{\ii\jj}-h_{\bm j\bm i}+\mu\delta_{\ii\jj}$. By definition the fields $\Sigma_{\ii\sigma}(\tau',\tau)$ are Hermitian,
\beg
\Sigma_{\ii\sigma}(\tau',\tau)=\Sigma_{\ii\sigma}^*(\tau,\tau').
\en
 The statement here is that integrating $e^{-S_\mathrm{eff}}$ with $S_\mathrm{eff}$ from \eref{S_Nambu} over the fields $\Phi_\ii(\tau',\tau)$ and $\Sigma_{\ii\sigma}(\tau',\tau)$, we obtain \eref{S_fermion}. On the stationary point, the fields  $\Phi_\ii(\tau',\tau)$ and $\Sigma_{\ii\sigma}(\tau',\tau)$ depend only on the difference $\tau'-\tau$ and represent the anomalous and normal self-energies (see below).

The action~\re{S_Nambu} is quadratic in fermion fields. Performing the Gaussian integral over these fields, we obtain the effective action
for the fields  $\Phi_\ii(\tau',\tau)$ and $\Sigma_{\ii\sigma}(\tau',\tau)$:
\beg
\begin{split}
 S_\mathrm{eff}  = 
\sum_\ii \int\!\!\!\!\int d\tau'd\tau \left[ \frac{ \Phi^*_\ii (\tau',\tau) \Phi_\ii(\tau',\tau)}{ D(\tau'-\tau) } +\sum_{\sigma}\frac{ \Sigma_{\ii\sigma}(\tau',\tau) \Sigma_{\ii\sigma}(\tau,\tau')}{2D(\tau'-\tau)}  \right] -\mathrm{Tr} \ln M 
 -\sum _{x'x} K^\dagger_{x'}   M^{-1}_{x'x}  K_{x}.
\end{split}
\label{S_fields}
\en
  It is convenient to transform the matrices and vectors in  $S_\mathrm{eff}$ to fermionic Matsubara frequencies $\omega_n$ and the eigenbasis of the single-electron Hamiltonian $\hat h$ with a unitary matrix
\beg
   U_{\tau n,\ii \alpha}=\sqrt{T}
\begin{bmatrix}
\pi_{\ii \alpha} e^{-i\tau\omega_n}  & 0\\
0 & \pi_{\ii \alpha} e^{-i\tau\omega_n}\\
\end{bmatrix}\!,
\en
which leaves $S_\mathrm{eff}$ invariant, since it is a scalar (matrix trace). Here $\pi_{\alpha}$ are the eigenstates of $\hat h$,
\beg
\sum_\jj h_{\ii \jj} \pi_{\jj \alpha} =\epsilon_\alpha \pi_{\ii\alpha}.
\en
For example, the matrix $M$ transforms into $(U^\dagger M U)_{nm,\alpha\beta}= \sum_{\ii\jj} \int\!\!\int d\tau'd\tau U^\dagger_{n\tau, \alpha\ii} M_{\ii\tau,\ii'\tau'}U_{\tau'm, \ii'\beta}$ and similarly $K\to U^\dagger K$ and $\Psi\to U^\dagger \Psi$. Carrying out this transformation in \eref{S_fields}, we find
\beg
\begin{split}
{ S}_\mathrm{eff}= T\!\!\!\!  \sum_{nml\alpha\beta} \left[ \left( \Phi^{\alpha\beta}_{n+l,m+l}\right)^*  D^{-1}_l  \Phi_{nm}^{\alpha\beta}+\frac{1}{2}\sum_\sigma 
\Sigma_{\sigma, m+l,n+l}^{\alpha\beta} D^{-1}_l \Sigma_{\sigma n m}^{\beta\alpha}\right]
-\Tr\!\ln M-  T K ^\dagger M^{-1}  K.
\end{split}
\label{seffab}
\en
Here $D_l^{-1}$ is the bosonic Matsubara frequency $\omega_l$ component of $1/D(\tau)$,
\beg
 \Phi_{nm}^{\alpha\beta}=T\sum_\ii \!\iint d\tau'd\tau e^{i\omega_n\tau'} \pi^*_{\ii\alpha}\Phi_\ii(\tau', \tau) \pi_{\ii\beta} e^{-i\omega_m\tau},
 \label{newold}
\en
and similarly for $\Sigma_{\sigma nm}$. Hermiticity of $\Sigma_{\ii\sigma}(\tau', \tau)$ implies Hermiticity of $\Sigma_{\sigma nm}^{\alpha\beta}$, since the transformation is unitary.
The matrix $M$ in the new basis reads as
\beg
M= 
\begin{bmatrix}
(-i\omega_n+\xi_\alpha)\delta_{nm}-i \Sigma_{\up nm}^{\alpha\beta} & \Phi_{nm}^{\alpha\beta}\\
\left(\Phi_{mn}^{\beta\alpha}\right)^* & (-i\omega_n-\xi_\alpha)\delta_{nm}+i \Sigma_{\dn, -m, -n}^{\alpha\beta}\\
\end{bmatrix},\\
\label{Malphabeta1}
\en
and the source fields are
\beg
\begin{split}
K_{\alpha m} =
\begin{bmatrix}
J_{\up \alpha m}\\
-J_{\dn, \alpha', -m}^*\\
\end{bmatrix},\quad 
  J_{\up \alpha m}=\sum_\ii \int d\tau e^{i\omega_m\tau} \pi_{\ii \alpha}^*J_{\ii \up}(\tau),\quad J_{\dn \alpha' m}=\sum_\ii \int d\tau e^{i\omega_m\tau} (\pi_{\ii \alpha}^*)^* J_{\ii\dn}(\tau).
\end{split}
\en
Note that $\alpha'$ labels the state $\pi_\alpha^*$, which is related to $\pi_\alpha$ by time-reversal operation and $A_{-m,-n}$ stands for $A(-\omega_m,-\omega_n)$.

\subsection{Stationary point}

The Eliashberg equations are a stationary point of the effective action, such that the fields $\Phi_\ii(\tau',\tau)$ and $\Sigma_{\ii\sigma}(\tau',\tau)$ are spatially uniform and depend only on the time difference,
\beg
\Phi_\ii(\tau',\tau)=\Phi(\tau'-\tau),\quad \Sigma_{\ii\sigma}(\tau'-\tau)=\Sigma_{\sigma}(\tau'-\tau).
\label{uniform}
\en
Equation~\re{newold} then implies
\beg
\Phi_{nm}^{\alpha\beta}=\Phi_{n} \delta_{nm}\delta_{\alpha\beta},\quad \Sigma_{\sigma nm}^{\alpha\beta}=\Sigma_{\sigma n} \delta_{nm}\delta_{\alpha\beta}.
\label{diag}
\en
In words, off-diagonal matrix elements are zero and diagonal ones depend only on the Matsubara frequency. 

The derivative of the action~\re{seffab} with respect to any off-diagonal matrix element is a sum of terms each of which is proportional to one of the other off-diagonal matrix elements. This means that setting these matrix elements to zero automatically solves all stationary point equations for them. To determine stationary point equations for the diagonal matrix elements, we substitute \eref{diag} into \eref{seffab}. Setting also the sources to zero, we obtain
\beg
\begin{split}
{ S}_\mathrm{eff}= T  N\sum_{nl}\left[ \Phi_{n+l}^* D_l^{-1} \Phi_{n} + \Sigma_{n+l} D_l^{-1}  \Sigma_{n} - \chi_{n+l} D_l^{-1} \chi_{n}  \right]
-\sum_{n\alpha} \ln \left[ (\omega_n+\Sigma_{n} )^2 +| \Phi_{n} |^2 + (\chi_{n} +\xi_\alpha)^2\right],
\end{split}
\label{seffdiagindep}
\en
where $N$ is the number of sites (which is equal to the number of the single-electron states $\pi_\alpha$) and
\beg
\Sigma_{n} =\frac{ \Sigma_{\up n}  - \Sigma_{\dn, -n} }{2},\quad i \chi_{n} =\frac{ \Sigma_{\up n} +\Sigma_{\dn, -n}  }{2},
\label{oddeven}
\en
 where $A_{-n}$ means $A(-\omega_n)$. 
Minimizing the effective action with respect to $\Phi_n^*, \Sigma_n$ and $\chi_n$, we derive three Eliashberg equations
\beg
\begin{split}
\Phi_n =  \frac{T}{N}\sum_{m\alpha} D_{n-m} \frac{\Phi_m }{(\omega_m+\Sigma_{m})^2 +| \Phi_{m} |^2 + (\chi_{m} +\xi_\alpha)^2},\\
\Sigma_n=  \frac{T}{N}\sum_{m\alpha} D_{n-m} \frac{\omega_m+\Sigma_m}{(\omega_m+\Sigma_{m})^2 +
| \Phi_{m}|^2 + (\chi_{m} +\xi_\alpha)^2},\\
\chi_n = - \frac{T}{N} \sum_{m\alpha} D_{n-m} \frac{\xi_\alpha+\chi_m }{(\omega_m+\Sigma_{m} )^2 +| \Phi_{m} |^2 + (\chi_{m} +\xi_\alpha)^2},
\end{split}
\label{elialleq}
\en
where
\beg
D_{n-m} =\frac{ \alpha^2 M^{-1}}{(\omega_n-\omega_m)^2+\Omega^2}.
\en
When  $\xi_\alpha$ are symmetric with respect to zero, $\chi_n\equiv0$ solves the last equation in~\re{elialleq}. 

As is normally done in the Migdal-Eliashberg theory, we now send the Fermi energy to infinity and take the density of states per site per spin orientation, $\nu_0$, to be constant. We implement this limit in \eref{seffdiagindep} by integrating the logarithm  over $\xi_\alpha$ from $-\mathbb{\Lambda}$ to $\mathbb{\Lambda}$,  discarding the constant term that depends only on $\mathbb{\Lambda}$, and then taking the limit $\mathbb{\Lambda}\to\infty$. The result is
\beg
\begin{split}
{ S}_\mathrm{eff}= T  N\sum_{nl}\left[ \Phi_{n+l}^* D_l^{-1} \Phi_{n} + \Sigma_{n+l} D_l^{-1}  \Sigma_{n}    \right]
-2\pi \nu_0 N \sum_{n} \sqrt{ (\omega_n+\Sigma_{n} )^2 +| \Phi_{n} |^2 }.
\end{split}
\label{seffsimple}
\en
Eliashberg equations become
\beg
\begin{split}
\Phi_n= \pi  T \sum_m \lambda(\omega_n-\omega_m)  \frac{ \Phi_m}{ \sqrt{(\omega_m+\Sigma_{m} )^2 +| \Phi_{m} |^2} },\\
\Sigma_n= \pi  T \sum_m \lambda(\omega_n-\omega_m)  \frac{\omega_m+ \Sigma_m}{ \sqrt{(\omega_m+\Sigma_{m} )^2 +| \Phi_{m} |^2}},
\end{split}
\label{elifamiliar1}
\en
with
\beg
\lambda(\omega_n-\omega_m)= \frac{  g^2}{(\omega_n-\omega_m)^2+\Omega^2},\quad g^2=  \nu_0\alpha^2 M^{-1}.
\en

The fermionic part of the free-energy functional per site is $f=TS_\mathrm{eff}/N$,
\beg
\begin{split}
f= \nu_0 T^2\sum_{nl}\left[ \Phi_{n+l}^* \Lambda_l  \Phi_{n} + \Sigma_{n+l} \Lambda_l   \Sigma_{n}    \right]
-2\pi \nu_0 T\sum_{n} \sqrt{ (\omega_n+\Sigma_{n} )^2 +| \Phi_{n} |^2 },
\end{split}
\label{feH}
\en
where $\Lambda_l$ is the discrete Fourier transform of $1/\lambda(\tau)$ at bosonic Matsubara frequency $\omega_l$.

\subsection{Spatially nonuniform stationary point}
\label{nonuapp}

For future reference let us also consider spatially nonuniform solutions
\beg
\Phi_\ii(\tau',\tau)=\Phi_\ii(\tau'-\tau),\quad \Sigma_{\ii\sigma}(\tau'-\tau)=\Sigma_{\ii\sigma}(\tau'-\tau).
\label{nuniform}
\en
Let $\Phi_{\ii n}$ be the Fourier transform  of $\Phi_\ii(\tau'-\tau)$ with respect to $\tau'-\tau$. Its
matrix elements in the eigenbasis   of the single-particle Hamiltonian $h_{\ii\jj}$  are $\Phi_n^{\alpha\beta}=\sum_\ii \pi^*_{\ii\alpha} \Phi_{\ii n} \pi_{\ii\beta}$. If the eigenstates $\pi_{\ii\alpha}$ are localized or delocalized but highly oscillatory, the off-diagonal, $\alpha\ne\beta$, matrix elements are negligible.  In this case,  a suitable ansatz  for the stationary point is
\beg
\Phi_{nm}^{\alpha\beta}=\Phi_{n}^\alpha \delta_{nm}\delta_{\alpha\beta},\quad \Sigma_{\sigma nm}^{\alpha\beta}=\Sigma_{\sigma n}^\alpha \delta_{nm}\delta_{\alpha\beta}.
\label{diag1}
\en
The derivation of the stationary point equations is similar to the uniform case. The effective action now reads as
\beg
\begin{split}
{ S}_\mathrm{eff}= T  \sum_{nl\alpha}\left[ (\Phi_{n+l}^\alpha)^* D_l^{-1} \Phi_{n}^\alpha + \Sigma_{n+l}^\alpha D_l^{-1}  \Sigma_{n}^\alpha - \chi_{n+l}^\alpha D_l^{-1} \chi_{n}^\alpha  \right]
-\sum_{n\alpha} \ln \left[ (\omega_n+\Sigma_{n}^\alpha )^2 +| \Phi_{n}^\alpha |^2 + (\chi_{n}^\alpha +\xi_\alpha)^2\right],
\end{split}
\label{seffdiagindepnu1}
\en
where  
\beg
\Sigma_{n}^\alpha =\frac{ \Sigma_{\up n}^\alpha  - \Sigma_{\dn, -n}^\alpha }{2},\quad i \chi_{n}^\alpha =\frac{ \Sigma_{\up n}^\alpha +\Sigma_{\dn, -n}^\alpha  }{2}.
\label{oddevennu1}
\en
It is incorrect to minimize ${ S}_\mathrm{eff}$ with respect to $\Phi_n^\alpha$, because $\Phi_n^\alpha=\sum_\ii |\pi_{\ii\alpha}|^2\Phi_{\ii n} $  are not independent. The original independent variables are $\Phi_{\ii n}$ and the matrix $R_{\ii\alpha}=|\pi_{\ii\alpha}|^2$ that
relates $\Phi_n^\alpha$ to $\Phi_{\ii n}$ is degenerate. For example, the completeness relation $\sum_\alpha \pi_{\ii\alpha} \pi_{\jj\alpha}=\delta_{\ii\jj}$ implies $\sum_\alpha R_{\ii\alpha}= 1$ meaning that the columns of the matrix $R$ are linearly dependent. Similarly, the normalization condition   $\sum_\ii |\pi_{\ii\alpha}|^2 =1$ says that the rows of $R$ are linearly dependent too. The same applies to  $\Sigma_n^\alpha$ and  $\chi_n^\alpha$.

 We should instead minimize \eref{seffdiagindepnu1} with respect to $\Phi_{\ii n}$, $\Sigma_{\ii n}$, and $\chi_{\ii n}$ using
 \beg
 \frac{\partial\Phi_n^\alpha}{\partial\Phi_{\ii n}} =|\pi_{\ii\alpha}|^2,\mbox{ etc.}
 \en
We obtain
\beg
\begin{split}
\sum_\alpha \Phi_n^\alpha |\pi_{\ii\alpha}|^2=  T \sum_{m\alpha} D_{n-m} \frac{\Phi_m^\alpha |\pi_{\ii\alpha}|^2}{(\omega_m+\Sigma_{m}^\alpha)^2 +| \Phi_{m}^\alpha |^2 + (\chi_{m}^\alpha +\xi_\alpha)^2},\\
\sum_\alpha \Sigma_n^\alpha |\pi_{\ii\alpha}|^2 =   T \sum_{m\alpha} D_{n-m} \frac{(\omega_m+\Sigma_m^\alpha) |\pi_{\ii\alpha}|^2}{(\omega_m+\Sigma_{m}^\alpha)^2 +
| \Phi_{m}^\alpha|^2 + (\chi_{m}^\alpha +\xi_\alpha)^2},\\
\sum_\alpha \chi_n^\alpha |\pi_{\ii\alpha}|^2 = -  T \sum_{m\alpha} D_{n-m} \frac{(\xi_\alpha+\chi_m^\alpha) |\pi_{\ii\alpha}|^2 }{(\omega_m+\Sigma_{m}^\alpha )^2 +| \Phi_{m}^\alpha |^2 + (\chi_{m}^\alpha +\xi_\alpha)^2}.
\end{split}
\label{elialleqnu}
\en
If the single-particle Hamiltonian $h_{\ii\jj}$ has the periodicity of the lattice, $\pi_{\ii\alpha}$ are plane waves and $|\pi_{\ii\alpha}|^2=N^{-1}$. Then,  
 $\Phi_n^\alpha=\Phi_n$, $\Sigma_n^\alpha=\Sigma_n$, $\chi_n^\alpha=\chi_n$, and we recover the usual Eliashberg equations~\re{elialleq}.

\subsection{Green's functions}
\label{gfapp}

Let us evaluate the Fourier transforms $\mathcal{G}_{\sigma\alpha n}$ and $\mathcal{F}_{\alpha n}$ of the normal and anomalous thermal Green's functions at the stationary point. The latter are defined as
\beg
\GG_{\sigma\alpha} (\tau-\tau')=- \langle T_\tau c_{\sigma\alpha}(\tau)  c^\dagger_{\sigma\alpha}(\tau')\rangle,\quad 
\FF_\alpha (\tau-\tau')= \langle T_\tau c_{\dn\alpha'}(\tau)  c_{\up\alpha}(\tau')\rangle.
\en
We obtain $\GG_{\sigma\alpha n}$ and $\FF_{\alpha n}$ by differentiating the partition function ${\cal Z}$ with respect to the source fields,
\beg
T \GG_{\sigma\alpha n} =\frac{1}{\cal Z} \frac{\partial^2 {\cal Z} }{\partial J_{\sigma\alpha n} 
\partial J^*_{\sigma\alpha n} },\quad 
 T \FF_{\alpha n} =\frac{1}{\cal Z} \frac{\partial^2 {\cal Z} }{ \partial J^*_{\dn\alpha' n} 
\partial J^*_{\up\alpha, -n} }.
\label{gfZ}
\en
The source fields enter through the $ K ^\dagger M^{-1}  K$ term in the effective action~\re{seffab}. At the stationary point this term becomes
\beg
\begin{split}
 K^\dagger M^{-1} K= \sum_{\alpha n} K_{n\alpha}^\dagger M^{-1}_{\alpha n} K_{n\alpha} =
   \sum_{\alpha n} \frac{1}{\Theta_{\alpha n} }\left[ \lefteqn{\phantom{\prod}}  a^+_{\alpha n} J^*_A J_A - \Phi_n J^*_A J^*_{A'} -
(\Phi_n)^*J_{A'} J_A +   a^-_{\alpha n} J_{A'} J^*_{A'} \right],
 \end{split}
\en
where $a^\pm_{\alpha n} =i(\omega_n+\Sigma_{n})\pm (\xi_\alpha +\chi_{n})$, the labels $A$ and $A'$ stand for $\up\!\!\alpha n$ and $\dn\!\alpha',\! -n$, respectively, and
$\Theta_{\alpha n} =(\omega_n+\Sigma_{n}^{\alpha})^2 +| \Phi_{n}^{\alpha}|^2 + (\chi_{n}^{\alpha}+\xi_\alpha)^2=-\det M_{\alpha n}$.

The source-dependent part of the partition function at the stationary point  is ${\cal Z}_s= e^{-T K^\dagger M^{-1} K}$.  Using \eref{gfZ}, we find
\beg
\begin{split}
\GG_{\up\alpha n}=-\frac{ i(\omega_n+\Sigma_{n} )+ \xi_\alpha +\chi_{n} } {(\omega_n+\Sigma_{n} )^2 +| \Phi_{n}|^2 + (\chi_{n} +\xi_\alpha)^2  },\\
\GG_{\dn\alpha n}=-\frac{i(\omega_n-\Sigma_{-n} )+ \xi_\alpha +\chi_{-n}   } { (\omega_n-\Sigma_{-n} )^2 +| \Phi_{-n} |^2 + (\chi_{-n} +\xi_\alpha)^2 },\\
\FF_{\alpha n}=-\frac{ \Phi_{-n}  } { (\omega_n-\Sigma_{-n} )^2 +| \Phi_{-n} |^2 + (\chi_{-n} +\xi_\alpha)^2}.
\end{split}
\label{gfanswer}
\en
Suppose there is a time reversal symmetry  so that $\Sigma_{\up n} =\Sigma_{\dn n} $ and $\GG_{\up\alpha n}=\GG_{\dn\alpha n}$. Equation~\re{oddeven} then implies that $\Sigma_n $ is odd in frequency and $\chi_n $ is even. It further follows from the above formulas for $\GG_{\up\alpha n}$ and $\GG_{\dn\alpha n}$ that $|\Phi_n|$ is even and 
 we arrive at the usual expressions for the thermal Green's functions in the Migdal-Eliashberg theory
\beg
\begin{split}
\GG_{\up\alpha n}=\GG_{\dn\alpha n}\equiv \GG_{\alpha n}=-\frac{ i(\omega_n+\Sigma_{n} )+ \xi_\alpha +\chi_{n} } {(\omega_n+\Sigma_{n} )^2 +| \Phi_{n} |^2 + (\chi_{n} +\xi_\alpha)^2  },\\
\FF_{\alpha n}=-\frac{ \Phi_{n}   } { (\omega_n+\Sigma_{n} )^2 +| \Phi_{n} |^2 + (\chi_{n} +\xi_\alpha)^2}.
\end{split}
\label{gfsym}
\en

\section{Path integral formulation of the Migdal-Eliashberg theory for dispersing phonons}
\label{standardapp}

Here we derive  Eliashberg equations and free energy for arbitrary phonon spectrum and momentum-dependent electron-phonon coupling. We will see that these  quantities  have the same form as for the Holstein model, but with    a more general kernel   which now involves an integral over the phonon spectrum. 

Our starting point is the standard electron-phonon Hamiltonian
\beg
\begin{split}
H=\sum_{\pp \sigma} \xi_\pp c^\dagger_{\pp\sigma} c_{\pp\sigma} +  \sum_\qq {\omega_{0}(\qq) } b^\dagger_\qq b_\qq+
\frac{1}{\sqrt{N}} \sum_{\pp \qq\sigma} \frac{\alpha_{\qq}}{\sqrt{2M\omega_0(\qq)}} c^\dagger_{\pp+\qq \sigma} c_{\pp\sigma} \left[ b^\dagger_{-\qq} + b_\qq\right].
\end{split}
\label{frol}
\en
Hermitian property requires $\alpha_\qq=\alpha_{-\qq}^*$. For simplicity, we take $\alpha_\qq$ and $\omega_0(\qq)$ to depend only on the magnitude  of the momentum $\qq$. Integrating out phonons, we obtain an effective action for the fermionic fields
\beg
\begin{split}
S_\mathrm{eff}=T\sum_{ \ps\sigma}  \psi^*_{\ps\sigma} G_{0\ps}^{-1} \psi_{\ps\sigma} 
- \frac{1}{2} \frac{T^3}{N} \sum_{ \ps\ps'\qs\sigma\sigma'} 
\frac{ |\alpha_q|^2 M^{-1}}{\omega_m^2+\omega_q^2} \psi^*_{ \ps+\qs\sigma} \psi_{ \ps\sigma}  \psi^*_{ \ps'\!-\qs\sigma'} \psi_{ \ps'\sigma'},
\end{split}
\label{dispseff}
\en
where $G_{0\ps}^{-1}= -i\omega_n+\xi_\pp$, and $\ps=(\omega_n, \pp)$; $\qs=(\omega_m, \qq)$ are the frequency-momentum 4-vectors. As in \eref{Ddefinition} we replace the bare phonon frequencies $\omega_0(q)$ with the renormalized frequencies $\omega_q$ 
in the denominator of the effective electron-electron interaction.

It is helpful to rewrite the interaction as
\beg
\begin{split}
\frac{1}{2}\sum_{\sigma\sigma'}\frac{T^2}{N}\frac{ |\alpha_q|^2 M^{-1}}{\omega_n^2+\omega_q^2} \psi^*_{ \ps+\qs\sigma} \psi_{ \ps\sigma}  \psi^*_{ \ps'-\qs\sigma'} \psi_{ \ps'\sigma'}=\psi^*_{\ps_1'\up} \psi^*_{-\ps_2'\dn} D^{\ps_1'\ps_2'}_{\ps_1\ps_2} \psi_{-\ps_2\dn} \psi_{\ps_1\up}-\\
-\frac{1}{2}\psi^*_{\ps_1'\up} \psi_{\ps_2'\up} D^{\ps_1'\ps_2'}_{\ps_1\ps_2} \psi^*_{\ps_2\up} \psi_{\ps_1\up}-
\frac{1}{2}\psi^*_{\ps_1'\dn} \psi_{\ps_2'\dn} D^{\ps_1'\ps_2'}_{\ps_1\ps_2} \psi^*_{\ps_2\dn} \psi_{\ps_1\dn},
\end{split}
\label{rewrite}
\en
where we treat $D$ as a matrix with matrix elements
\beg
D^{\ps_1'\ps_2'}_{\ps_1\ps_2}=D_\qs\delta_{\ps_1', \ps_1+q}\delta_{\ps_2', \ps_2+\qs},\quad D_\qs=\frac{T^2}{N}\frac{ |\alpha_q|^2 M^{-1}}{\omega_n^2+\omega_q^2}.
\en
As before, we decouple the interaction with fields $\Phi$ and $\Sigma_\sigma$:
\beg
\begin{split}
S_\mathrm{eff}=  T\!\!\!\!\! \sum_{\ps_1'\ps_2'\ps_1\ps_2} \left\{\Phi_{\ps_1'\ps_2'}^* [D^{-1}]^{\ps_1'\ps_2'}_{\ps_1\ps_2}   \Phi_{\ps_1\ps_2}+\frac{1}{2}\sum_\sigma\Sigma_{\sigma \ps_1'\ps_2'}^* [D^{-1}]^{\ps_1'\ps_2'}_{\ps_1\ps_2} 
\Sigma_{\sigma p_1p_2}\right\}+
  T\sum_{\ps_1\ps_2}\Psi_{\ps_1}^\dagger M_{\ps_1\ps_2}\Psi_{\ps_2},
\end{split}
\en
where
\beg
M_{\ps_1\ps_2}=
\begin{pmatrix}
[-i\omega_n+\xi_{\pp_1}] \delta_{\ps_1\ps_2} - i\Sigma_{\up \ps_1\ps_2} & \Phi_{\ps_1\ps_2} \\
 \Phi_{\ps_2\ps_1}^* & [-i\omega_n-\xi_{\pp_1}]  \delta_{\ps_1\ps_2} + i\Sigma_{\dn, -\ps_2,-\ps_1}\\
 \end{pmatrix},
 \quad
 \Psi_p=\begin{pmatrix}
\psi_{\ps\up }\\
 \psi^*_{-\ps\dn }\\
\end{pmatrix}.
 \en
Integrating out the fermions, we find
\beg
S_\mathrm{eff}=  T\!\!\!\!\! \sum_{\ps_1'\ps_2'\ps_1\ps_2} \left\{\Phi_{\ps_1'\ps_2'}^* [D^{-1}]^{\ps_1'\ps_2'}_{\ps_1\ps_2}   \Phi_{\ps_1\ps_2}+\frac{1}{2}\sum_\sigma\Sigma_{\sigma \ps_1'\ps_2'}^* [D^{-1}]^{\ps_1'\ps_2'}_{\ps_1\ps_2} 
\Sigma_{\sigma p_1p_2}\right\}-\Tr\!\ln M.
\label{actmomfull}
\en

\subsection{Stationary point}

It is reasonable to expect  that in a translationally invariant system, the stationary point is also translationally invariant.  Translational invariance
means that matrices $\Phi_{\ps_1\ps_2}$ and $\Sigma_{\sigma \ps_1\ps_2}$ are diagonal, 
\beg
\Phi_{\ps_1\ps_2}=\Phi_{\ps_1}\delta_{\ps_1\ps_2},\quad \Sigma_{\sigma \ps_1 \ps_2}=\Sigma_{\sigma \ps_1}\delta_{\ps_1\ps_2},
\en
since $\ps_1=\ps_2$ ensures that $e^{i \ps_1 x_1-i\ps_2 x_2}=e^{i \ps_1(x_1-x_2)}$ depends only on $\rr_1-\rr_2$ and $\tau_1-\tau_2$.

Setting the off-diagonal matrix elements to zero in the effective action~\re{actmomfull}, we obtain
\beg
\begin{split}
{ S}_\mathrm{eff} = T\sum_{\ps' \ps}\left\{ \Phi_{\ps'}^* [D^{-1}]^{\ps'}_{\ps}   \Phi_{\ps}+  \Sigma_{  \ps'} [D^{-1}]^{\ps'}_{\ps} 
\Sigma_{ \ps}  - \chi_{  \ps'} [D^{-1}]^{\ps'}_{\ps} 
\chi_{ \ps}\right\}
-\sum_\ps \ln \left[ (\omega_n+\Sigma_\ps)^2 +| \Phi_p|^2 + (\chi_\ps+\xi_\pp)^2\right], 
\end{split}
\label{seffdiag1}
\en
where
 \beg
\Sigma_{\ps} =\frac{ \Sigma_{\up \ps} - \Sigma_{\dn, -\ps}}{2},\quad i\chi_{\ps} =\frac{ \Sigma_{\up \ps} +\Sigma_{\dn, -\ps}  }{2}.
\label{oddeven1}
\en
Now  we evaluate the stationary point of this effective action  with respect to $\Phi_\ps^*$, $\Sigma_\ps$, and $\chi_\ps$ and multiply the resulting expression by matrix $D$
on both sides. The result is
\beg
\begin{split}
\Phi_{n \pp}= \frac{T}{N} \sum_{m\qq} \frac{ |\alpha_q|^2 M^{-1}}{(\omega_n-\omega_m)^2+\omega_{q}^2} \frac{\Phi_{m \pp'} }{(\omega_{m}+\Sigma_{m\pp'})^2 +| \Phi_{m\pp'}|^2 + (\chi_{m\pp'}+\xi_{\pp'})^2 },\\
\Sigma_{n\pp}= \frac{T}{N} \sum_{m\qq} \frac{ |\alpha_q|^2 M^{-1}}{(\omega_{n}-\omega_m)^2+\omega_{q}^2} \frac{  \omega_{m}+\Sigma_{m\pp'} }{(\omega_{m}+\Sigma_{m\pp'})^2 +| \Phi_{m\pp'}|^2 + (\chi_{m\pp'}+\xi_{\pp'})^2 },\\
\chi_{n\pp}=- \frac{T}{N} \sum_{m\qq} \frac{ |\alpha_q|^2 M^{-1}}{(\omega_{n}-\omega_m)^2+\omega_{q}^2} \frac{ \chi_{m\pp'}+\xi_{\pp'} }{(\omega_{m}+\Sigma_{m\pp'})^2 +| \Phi_{m\pp'}|^2 + (\chi_{m\pp'}+\xi_{\pp'})^2 },
\end{split}
\label{momall}
\en
where $\pp'=\pp+\qq$.  These are the momentum-dependent Eliashberg equations for the Hamiltonian~\re{frol} [cf. \eref{elialleq} for the Holstein model].

We look for an isotropic solution of \eref{momall}, so that $\Phi_{n \pp}$, $\Sigma_{n\pp}$, and $\chi_{n\pp}$ are independent of the direction of $\pp$. Then,  it is useful to  convert the summation over $\qq$ in \eref{momall} into an integral over $q$ and $p'$, where $q$ and $p'$ are the magnitudes of vectors $\bm q$ and $\bm p'$. Using ${p'}^2=p^2+q^2+2pq\cos\theta$, we evaluate the Jacobian for the change of variables $(q, \cos\theta)\to (q, p')$ and find
\beg
\sum_\qq=\frac{V}{4\pi^2} \iint q^2 dq d\cos\theta = \frac{V}{4\pi^2}  \frac{1}{p} \int qdq \int p' dp'.
\label{mes1}
\en
Since the right hand side of \eref{momall} is a product of a function of $q$ only and a function of $p'$ only, the integration over $\qq$ factorizes into  a product of an integral  over $q$ and an integral over $p'$. 

In the Migdal-Eliashberg theory, the fluctuations around the stationary point are negligible only when the Fermi energy is much larger than any other characteristic energy~\cite{migdal,eli1st}. This means that the integration is confined to the vicinity of the Fermi surface, which we take to be spherical, and $p\approx p'\approx p_F$, where $p_F$ is the Fermi momentum. Since $\qq$ connects $\pp$ and $\pp'$, which are both on the Fermi surface, its magnitude varies between 0 and $2p_F$. This allows as to rewrite \eref{mes1} as
\beg
\sum_\qq=  \frac{V}{4\pi^2}  \frac{1}{p_F^2} \int_0^{2p_F}\!\!\!\! qdq \int {p'}^2 dp'= \frac{1}{2p_F^2} \int_0^{2p_F}\!\!\!\!  qdq \sum_{\pp'}.
\label{mes}
\en
Using this in \eref{momall}, we observe $\Phi_{n \pp}$, $\Sigma_{n\pp}$, and $\chi_{n\pp}$ are independent of the momentum $\pp$ and
performing the summation over $\pp'$ we arrive at the following two equations:
\beg
\begin{split}
\Phi_n= \pi  T \sum_m  \lambda(\omega_n-\omega_m) \frac{ \Phi_m}{ \sqrt{(\omega_m+\Sigma_{m} )^2 +| \Phi_{m} |^2} },\\
\Sigma_n= \pi  T \sum_m  \lambda(\omega_n-\omega_m)  \frac{\omega_m+ \Sigma_m}{ \sqrt{(\omega_m+\Sigma_{m} )^2 +| \Phi_{m} |^2}},
\end{split}
\label{elifamiliar2}
\en
where
\beg
 \lambda(\omega_l)=\frac{1}{2 p_F^2} \int_0^{2p_F}   \frac{g_q^2 q dq}{\omega_l^2+\omega_{q}^2},\quad g_q^2= |\alpha_q|^2 M^{-1}.
 \label{1morelam}
 \en
 This expression   is for a spherical Fermi surface in $d=3$ dimensions, but it is straightforward to extend it to any   $d\ge 2$. 
 Equations~\re{elifamiliar2} are of the same form as \esref{elifamiliar1}, but with a more general kernel $\lambda(\omega_n-\omega_m)$. For dispersionless phonons, 
$\omega_q=\Omega$  and $g_q=g$, this kernel too coincides with that in \eref{elifamiliar1}. By definition the dimensionless coupling constant for
dispersing phonons is
\beg
 \lambda= \lambda(\omega_l=0)= \frac{1}{2p_F^2}  \int_0^{2p_F}  \frac{  g_q^2 q dq}{\omega^2_{q}}.
 \en
 
 The effective action evaluated at the stationary point is [cf. \eref{seffsimple}]
\beg
\begin{split}
{ S}_\mathrm{eff}= T \nu_0 V\sum_{nl}\left[ \Phi_{n+l}^* \Lambda_l  \Phi_{n} + \Sigma_{n+l} \Lambda_l   \Sigma_{n}    \right]
-2\pi \nu_0 V\sum_{n} \sqrt{ (\omega_n+\Sigma_{n} )^2 +| \Phi_{n} |^2 }.
\end{split}
\label{seffsimple1}
\en
Here $\Lambda_l $ is the Fourier transform of $1/\lambda(\tau)$ at frequency $\omega_l$. The corresponding free energy $f=TS_\mathrm{eff}/N$ per lattice site  is
\beg
\begin{split}
f=  \nu_0 T^2 \sum_{nl}\left[ \Phi_{n+l}^* \lambda_l^{-1} \Phi_{n} + \Sigma_{n+l} \lambda_l^{-1}  \Sigma_{n}    \right]
-2\pi \nu_0 T\sum_{n} \sqrt{ (\omega_n+\Sigma_{n} )^2 +| \Phi_{n} |^2 }.
\end{split}
\label{fe}
\en

\subsection{BCS theory as the weak coupling limit of the Migdal-Eliashberg theory}
\label{weak}

Confusion exists in the literature to this day regarding whether or not the BCS theory is the weak coupling limit of the Migdal-Eliashberg theory~\cite{mars1,mars2}. This confusion arises from attempts to compare cutoff-dependent quantities which are illegitimate within the BCS theory. The BCS theory is
only valid in the limit $\Delta_\mathrm{BCS}/\Xi\to0$, where $\Delta_\mathrm{BCS}$ is the BCS ground state gap and $\Xi$ is the cutoff of the order of
the Debye energy   for acoustic phonons and $\Omega_0$ for Einstein phonons (Holstein model). 

In other words, one should take the limit
$\Xi\to\infty$ and the coupling constant $\lam\to 0$ while keeping $\Delta_\mathrm{BCS}\propto\Xi e^{-1/\lam}$ fixed. Only quantities that  survive this limit are  legitimate within the BCS theory. In particular, it is meaningless to compare $\Delta_\mathrm{BCS}/\Xi$ or $T_c/\Xi$, but, for example, $\Delta_\mathrm{BCS}/T_c$, the condensation energy, the normalized jump in the specific heat are meaningful and their BCS values should agree with those in the weak coupling limit of the Migdal-Eliashberg theory, which they indeed do~\cite{bardeen,mitr,marsiglio,carbotte}.

It is   well-understood how the BCS theory emerges from the Migdal-Eliashberg theory in the weak coupling limit. Here we reproduce known arguments within our path integral framework. Our aim in doing so is to be able to compare later the weak and strong coupling limits of the Migdal-Eliashberg theory as well as Anderson and Eliashberg spins.  
 We define the weak coupling limit as the limit where the ratio of the energy gap to the characteristic phonon frequency $\omega_\mathrm{ch}$ goes to zero. This is equivalent to $\omega_\mathrm{ch} \to\infty$ (cf. $\omega_\mathrm{ch}=\Omega\to 0$ in the strong coupling limit of the Holstein model [see~\eref{lam}]). 
 
 Since  relevant energies and frequencies are of the order of the gap, $\omega_n$ in the denominator is negligible in the effective action~\re{dispseff}. Then, the interaction is instantaneous and the effective action corresponds to the Hamiltonian
\beg
H=\sum_{\pp\sigma} \xi_\pp c^\dagger_{\pp\sigma} c_{\pp\sigma} -\frac{1}{2N} \sum_{\pp\pp'\qq\sigma\sigma'} \frac{ \alpha_q^2 M^{-1}}{\omega_q^2} c^\dagger_{\pp+\qq\sigma} c_{\pp\sigma} c^\dagger_{\pp'-\qq\sigma'}c_{\pp'\sigma'}. 
\label{almostbcs}
\en
The hierarchy of scales  $\eps_F\gg \omega_\mathrm{ch}\gg\Delta_\mathrm{BCS}$ implies that the pairing interaction is confined to the vicinity of the Fermi surface and that 
  momenta of phonons mediating the interaction are of the order of the Fermi momentum. Therefore, the range of the electron-electron interaction is of order of the Fermi wavelength $\lam_F$.  The characteristic length scale (coherence length)  in a weakly coupled superconductor  is of the order of $v_F/\Delta_\mathrm{BCS}$, which is orders of magnitude  larger than $\lam_F$. We see that the dimensionless range of  interactions goes to zero in the weak coupling limit. This means the interaction potential is a delta function in the position space and a constant in the momentum space. Since the pairing occurs at the Fermi surface in this limit, we determine the constant by averaging the interaction over this surface,
\beg
 \mathrm{const}=\frac{1}{2 p_F^2}  \int_0^{2p_F}  \frac{ \alpha_q^2 M^{-1} q   dq}{\omega_{q}^2}=\frac{\lam}{\nu_0},
 \en
 where we used~\eref{lam}. This is very similar to how we obtained~\eref{1morelam} in the Migdal-Eliashberg theory. The interaction in the position basis takes the form $c^\dagger_{\sigma}(\rr) c_{\sigma}(\rr) c^\dagger_{\sigma'}(\rr)c_{\sigma'}(\rr)$ For $\sigma=\sigma'$ this is equal to $c^\dagger_{\sigma}(\rr) c_{\sigma}(\rr)$, which we absorb into the chemical potential term.
 We are left with $\sigma\ne\sigma'$ terms, i.e., with the BCS Hamiltonian
\beg
H=\sum_{\pp\sigma} \xi_\pp c^\dagger_{\pp\sigma} c_{\pp\sigma} - \lam\delta \sum_{\pp\pp'\qq}   c^\dagger_{\pp+\qq\up}  c^\dagger_{\pp'-\qq\dn}c_{\pp'\dn} c_{\pp\up},
\label{bcsnotred}
\en
where $\delta=(\nu_0 N)^{-1}$ is the single-particle level spacing at the Fermi energy.

Both BCS and Eliashberg theories are mean-field theories. In mean-field approach, only Cooper pairs with zero total momentum determine the ground state and low-lying excitation spectrum. Keeping only such pairs, we end up with the reduced BCS Hamiltonian~\cite{bcs}
\beg
H=\sum_{\pp\sigma} \xi_\pp c^\dagger_{\pp\sigma} c_{\pp\sigma}-\lam\delta \sum_{\pp\pp'} c^\dagger_{-\pp\up} c^\dagger_{\pp\dn} c_{\pp'\dn} c_{-\pp'\up}.
\label{bcsred}
\en

We expect the above sequence of steps leading to~\eref{bcsred} to be exact when we take both the weak coupling and thermodynamic limits. The neglect of frequency and momentum dependencies of the electron-electron interaction is valid at energies much smaller than $\omega_\mathrm{ch}$. It resulted in infinite-range interaction in momentum space between Cooper pairs in~\eref{bcsred} [or, equivalently, spins
in~\eref{bcss}]. Therefore, this infinite-range interaction has to be cut off at $\Xi\sim\omega_\mathrm{ch}$. Any quantity to which energies
of order $\omega_\mathrm{ch}$ contribute is beyond the BCS theory.

The weak coupling limit of the Migdal-Eliashberg theory, the BCS theory, is universal in the sense that it is independent of the phonon dispersion and momentum dependence of the electron-phonon coupling. There is only one low-energy scale in this limit: the BCS gap $\Delta_\mathrm{BCS}$. The gap itself \textit{cannot} be determined accurately from within the BCS theory. However, the ratio of any other energy to the gap, e.g., $T_c/\Delta_\mathrm{BCS}$ is well defined.  The strong coupling limit of the Migdal-Eliashberg theory is similarly universal~\cite{combescot}.


\begin{thebibliography}{99}
\bibitem{migdalthm} This theorem says that corrections to the Migdal-Eliashberg theory are negligible when $\lam_0 \omega_\mathrm{ch}/\eps_F\to0$, where $\eps_F$ is the Fermi energy, \textit{and} when $\lam_0$  is sufficiently small~\cite{migdal,eli1st}, namely, ``not too close too $\frac{1}{2}$''~\cite{agd}.
\bibitem{arbitrary} The arbitrariness of $\hat h$ does not play an essential role in this paper, but will be important later~\cite{breakEli}.
\bibitem{grand} We often refer to $f$ as the free energy or  free-energy density. To be more precise,  taken at its minimum $f$ is the  grand potential per site as we work at fixed chemical potential.
\bibitem{zeta} This constant is in fact zero by the zeta function regularization prescription~\cite{kleinert}.  Indeed, the summand here depends only on $p=n-m$. Summation over $p$ leaves us with a sum over $l=n+m$ of the form $\sum_{l=-\infty}^\infty 1= 1+2\zeta(0)=0$.  See also Ref.~\cite{breakEli}.
\bibitem{trueT0} As usual, $T=0$ on Matsubara axis is understood as the limit $T\to0$ and
 we assume $\Delta(\omega)$ [$T\to0$ limit of $\Delta(\omega_n)$] is at least piecewise continuous. It is clear then that all these properties commute with $T\to0$ limit.
\bibitem{subtract} One must subtract a constant from $H_s$, replacing $\omega_n\cos\theta_n$ with $\omega_n\cos\theta_n-|\omega_n|$ and
$\cos(\theta_n-\theta_m)$ with $\cos(\theta_n-\theta_m)-1$, to make the summations converge. The first replacements subtracts the free energy of noninteracting Fermi gas, the second is discussed below \eref{spinh21}. This does not affect the stationary points in any way.
\bibitem{note234} These papers use the same expression for the free energy as we do (see Eq.~(39) in Ref.~\cite{andreygamma2}), except they subtract the normal state free energy and accordingly call it the condensation energy. They also take $\lam_{nm}=g^2/(\omega_n-\omega_m)^\gamma$ and consider
other values of $\gamma$ in addition to $\gamma=2$ (our case). However, all results of this section hold equally well for all values of $\gamma>0$.
\bibitem{high} The high energy scale is the Fermi energy $\eps_F$. The hierarchy of scales in the strong coupling limit is $\eps_F\gg g\gg \Omega$ and must be maintained when $\lam\to\infty$. The analogous hierarchy in the weak coupling limit is $\eps_F\gg \Omega\gg \Delta_\mathrm{BCS}$.
\bibitem{morep} More precisely, they correspond to those eigenstates that do not contain unpaired fermions, i.e., to the BCS ground state and excitations of the BCS condensate.
\end{thebibliography}
\end{document}